\documentclass[12pt,a4paper]{article}
\usepackage{amsmath}
\usepackage{multirow,booktabs}
\usepackage{graphicx}
\usepackage{amssymb}
\usepackage{amsthm}
\usepackage{enumitem}
\usepackage{color}
\usepackage{setspace} 

\allowdisplaybreaks

\begin{document}
	\title{\sc Order Restricted Inference for Adaptive Progressively Censored Competing Risks Data}
	\date{}
	
   \author{Ayon Ganguly \footnote{Department of Mathematics, Indian Institute
   of Technology Guwahati, Guwahati 781039, Assam, India. Email:
   aganguly@iitg.ac.in}, Debanjan Mitra \footnote{Operations Management,
   Quantitative Methods and Information Systems Area, Indian Institute of
   Management Udaipur, Udaipur 313001, Rajasthan, India. Email:
   debanjan.mitra@iimu.ac.in}, \& Debasis Kundu\footnote{Department of
   Mathematics and Statistics, Indian Institute of Technology Kanpur, Kanpur
   208016, Uttar Pradesh, India. Email: kundu@iitk.ac.in}}
	
	\date{}
	\maketitle

	\begin{abstract}
       Under adaptive progressive Type-II censoring schemes, order restricted inference
       based on competing risks data is discussed in this article. The latent failure
       lifetimes for the competing causes are assumed to follow Weibull distributions,
       with an order restriction on the scale parameters of the distributions. The
       practical implication of this order restriction is that one of the risk factors is
       dominant, as often observed in competing risks scenarios. In this setting,
       likelihood estimation for the model parameters, along with bootstrap based
       techniques for constructing asymptotic confidence intervals are presented. Bayesian
       inferential methods for obtaining point estimates and credible intervals for
       the model parameters are also discussed. Through a detailed Monte Carlo simulation
       study, the performance of order restricted inferential methods are assessed. In
       addition, the results are also compared with the case when no order restriction is
       imposed on the estimation approach. The simulation study shows that order
       restricted inference is more efficient between the two, when this additional
       information is taken into consideration. A numerical example is provided for
       illustrative purpose.
	\end{abstract}
	
	
   
   \newpage

	\doublespacing
\section{\sc Introduction}\label{sec:Intro}
Reliability researchers have given significant attention to the analysis of Type-II progressively censored
data which are obtained from a scheme as follows.
Suppose $n$ units are placed on a life testing experiment, and the number of failures to
be observed, say $m$, is fixed. 
Let 
\begin{equation}
X_{1:m:n} < X_{2:m:n} < ... < X_{m:m:n}   \label{ord-data}
\end{equation}
denote the ordered observed failure times. Immediately after the first observed failure at
$X_{1:m:m}$, $R_1$ functioning units are randomly removed from the experiment. Similarly,
immediately after $X_{2:m:n}$, $R_2$ functioning units are randomly removed from the
experiment, and so on. At the time of the $m-$th failure, all the remaining functioning units are
removed, and the experiment is terminated. The censoring scheme, given by $\boldsymbol R =
(R_1, R_2, ..., R_m)$, is pre-specified, and naturally, $m+\sum_{i=1}^mR_i = n$.
Progressive Type-II censoring is a general version of the conventional Type-II censoring
scheme, as can be easily seen by setting $(R_1, R_2, ..., R_m)$ to $(0, 0, ..., 0, n-m)$.
The progressive censoring scheme has many desirable properties, including that it tends to
give a more detailed picture of the tail behaviour of the underlying lifetime distribution.
For details on this topic, refer to the excellent account by Balakrishnan and
Cramer~\cite{B:BC2014}. 

The progressive Type-II censoring scheme has a longer test duration compared to the
conventional Type-II censoring scheme in return to the efficiency of inference
(see Burkschat~\cite{B2008}, Ng et al.~\cite{NCB2004}). The adaptive progressive Type-II
censoring scheme, proposed by Ng et al.~\cite{NKC2009}, introduced a controlling parameter
$T$ that controls the total duration of a lifetime experiment. At the beginning of the
experiment, along with the progressive censoring scheme $(R_1, R_2, ..., R_m)$, the
experimenter provides a time $T$ as the ideal duration of the
experiment. If $m$ failures are obtained before $T$, the censoring is carried out
according to the pre-specified scheme.  However, if less than $m$ failures are observed
till $T$, then the censoring scheme is modified, with the goal of terminating the
experiment as soon as possible after $T$. The adaptive progressive Type-II censoring
scheme is thus a very useful method to be used in practical reliability experiments as it controls
the total test duration while retaining the desirable properties of progressive censoring. 

In competing risks scenarios, there could be multiple risk factors that may
cause the failure of a unit; a researcher may want to assess one or more of the risk
factors in particular. Competing risks have been studied extensively in reliability
literature; see for example Pascual~\cite{P2007}, Pascual~\cite{P2008}, Pareek et
al.~\cite{PKK2009} and the references therein. The statistical inferential
issues for adaptive progressively censored data in the presence of competing risks were
addressed by Ren and Gui~\cite{RG2021}. The authors assumed the latent failure time model,
where each latent failure time had a Weibull distribution. They considered likelihood and
Bayesian inferences.

It is common for a researcher to have an additional information that one of the risk
factors is more severe compared to the other. A possible way to incorporate this
information into the statistical model is to consider an order restriction on the suitable
parameters of the underlying life distributions. Of course, in absence of such an
information, no assumption of order restriction among parameters is required. The order
restricted inferences are considered by several authors in different context; for example,
the readers are referred to Balakrishnan et al.~\cite{BBK2009}, Samanta et
al.~\cite{SGKM2017}, Pal et al.~\cite{AMK2021a, AMK2021b}, and Mahto et
al.~\cite{MLTW2021}. Balakrishnan et al.~\cite{BBK2009} used isotonic regression technique
to obtain MLEs of the model parameters under step-stress setup. It may be noted that the
implementation of the isotonic regression is quite complicated, which can be seen from (9)
of the article. Therefore, in this article, we reparameterize the original model
parameters to incorporate order restriction on original model parameters, as proposed by
Samanta et al.~\cite{SGKM2017}.

In this paper, our main aim is to consider inference under an order restriction on the
scale parameters of the life distributions for adaptive progressive Type-II censored
competing risks data.  Likelihood as well as Bayesian inferential methods are discussed.
The proposed methods of inference are then assessed through a detailed Monte Carlo
simulation study. We study the effect of order restriction on parameter estimates by
comparing order restricted inference with unrestricted inference through extensive
simulation. It is observed that if the true value of the model parameters are close, the
estimates of some parameters obtained using order restricted inference have higher
precision compared to the estimates obtained when there is no ordering assumed on the
parameters.

The rest of the paper is organized as follows. In Section~\ref{sec:Structure}, the
structure of adaptive progressively Type-II censored competing risks data, and the
notations used are presented. In Section~\ref{sec:InfOrder}, we discuss likelihood and
Bayesian inferential methods for the model parameters under an order restriction on the
scale parameters. The inferential results without the order restriction are briefly stated
in Section~\ref{sec:InfNoOrder}.
Section~\ref{sec:Simu} presents results of a detailed Monte Carlo simulation study in
which we assess performance of all the inferential procedures developed under order
restriction in Section~\ref{sec:InfOrder}, and compare them with the case when there is no
order restriction. A data analysis is provided in Section~\ref{sec:Illus} for illustrative
purpose. Finally, the paper is concluded in Section~\ref{sec:Con} with some remarks.

\section{\sc Structure of the data and notations used}\label{sec:Structure}
Suppose $n$ items are put on a life test, and the researcher wants to observe $m$ failures.
Let the ordered observed failure times be denoted by (\ref{ord-data}). At the beginning of
the experiment, the experimenter provides a censoring scheme $\boldsymbol R = (R_1, R_2,
..., R_m)$ with $m+\sum_{i=1}^mR_i = n$, and the ideal duration of the life test $T$.
Then, two scenarios may arise:

\noindent {\sc Case I:} $X_{1:m:n} < X_{2:m:n} < ... < X_{m:m:n} < T$. In this case, the experiment stops at $X_{m:m:n}$, and the random removal of units are carried out according to the pre-specified censoring scheme $(R_1, R_2, ..., R_m)$. \newline
\noindent {\sc Case II:} For some $J = 0, 1, ..., m-1$, $X_{J:m:n} < T < X_{J+1:m:n}$,
with $X_{0:m:n}=0$. Once the experimental time exceeds $T$ but the number of observed
failures has not reached $m$, the experimenter would want to terminate the experiment as
soon as possible. Now, it is known from the theory of order statistics that larger the
number of operating units left on test, the smaller the expected total test time
(see Ng et al.~\cite{NKC2009}, David and Nagaraja~\cite{B:DN2003}). In view of this, the
experimenter would leave as many items as possible on test, in order to not go too far
from the ideal test duration $T$.  Therefore, according to the proposed method by Ng et
al.~\cite{NKC2009}, the censoring scheme would be modified as 
$$R_{J+1} = .... = R_{m-1} = 0, \quad R_m = n - m - \sum_{i=1}^JR_i.$$
Thus, under adaptive progressive Type-II censoring, the scheme becomes
\begin{eqnarray}
R_j^* =
\begin{cases}
R_j, & \textrm{for} \quad j = 1,2,...,J \\ \nonumber
0, & \textrm{for} \quad j= J+1,J+2,...,m-1 \\ \nonumber
n-m-\sum_{j=1}^JR_j, & \textrm{for} \quad j=m.
\end{cases}
\end{eqnarray}
Note that Case II includes Case I when $J = m$, with $X_{m+1:m:n} = \infty$.  

The value of $T$ plays a very important role to determine the censoring scheme $(R_1, ...,
R_m)$. In particular, as per requirement, $T$ can be tuned to have a shorter experimental
time, or to have a higher probability of observing large lifetimes. When $T \to \infty$,
the scheme is the conventional progressive Type-II censoring, and when $T \to 0$, it is
the conventional Type-II censoring. Although, choosing an optimal $T$ is an important
issues, it is not pursued here.

To model competing risks data, there are two possible approaches - the latent failure time
modelling approach (Cox~\cite{C1959}) and the cause-specific hazard modelling approach
(Prentice et al.~\cite{PKPFFB1978}). However, as Kundu~\cite{K2004} observed, when the
underlying lifetime distribution is exponential or Weibull, the two approaches lead to the
same likelihood, though the interpretation of different probabilities under these two
approaches may be different. In this paper, we use the latent failure time modelling
approach of Cox~\cite{C1959}. We assume that the lifetimes under each competing risk
factors follow a Weibull distribution. 

Here are the notations we use in this paper: \newline
\noindent $X_i$: lifetime under cause $i$, $i$=1,2 \newline
\noindent $X$: observed lifetime, i.e., Min$\{X_1, X_2\}$  \newline
\noindent $T$: ideal duration of the test \newline
\noindent $\boldsymbol R = (R_1, ..., R_m)$: pre-specified progressive censoring scheme \newline
\noindent $\boldsymbol R^* = (R_1^*, ..., R_m^*)$: adaptive progressive censoring scheme \newline
\noindent $\delta$: indicator variable to indicate the type of failure (1 if failure is from cause 1; 2 if failure is from cause 2) \newline
\noindent $I_i$: index set of failures from cause $i$, $i$=1, 2, i.e., $I_i = \{j: \delta_j=i\}$ \newline
\noindent $|I_i|$: cardinality of $I_i$; We assume that $|I_i| = m_i$, $i$=1, 2 and $m = m_1+m_2$ \newline
\noindent We$(\alpha, \lambda)$: Weibull distribution with probability density function $\alpha \lambda x^{\alpha - 1}e^{-\lambda x^{\alpha}}$; $x>0$. 

We assume that $X_1$ and $X_2$ are independently distributed Weibull random variables with a common shape parameter, and different scale parameters, i.e., $X_1 \sim$ We$(\alpha, \lambda_1)$, and $X_2 \sim$ We$(\alpha, \lambda_2)$.

\section{\sc Order restricted Inference}\label{sec:InfOrder}
\subsection{\sc Likelihood inference}\label{subsec:LikInfOrder}
It is common to encounter situations where one of the competing risk factors is more
dominating compared to the other. In these cases, one can expect to observe more failures
from this dominating risk factor, compared to the other risk factor. To incorporate this
information into the model, we can impose an order restriction on the scale parameters of
the assumed distributions. That is, we can assume that $\lambda_1 > \lambda_2$, and
develop inferences with this order restriction. 

Let $X_1$ and $X_2$ denote the lifetimes corresponding to the two risk factors that
independently follow We$(\alpha, \lambda_1)$ and We$(\alpha, \lambda_2)$, respectively.
Further, suppose $\lambda_2 = \beta \lambda_1$, where $0 < \beta \le 1$. We develop
inferences under Case-II, as it includes Case-I. The likelihood function in Case-II is  
\begin{eqnarray}
& L(\boldsymbol \theta | Data) \propto & \alpha^{m}\lambda_1^{m}\beta^{m_2} \prod_{i=1}^{m}x_{i:m:n}^{\alpha-1} \times e^{-\lambda_1(1+\beta)\sum_{i=1}^m(R_i^*+1)x_{i:m:n}^{\alpha}}, \label{order-lik}
\end{eqnarray}
where $\boldsymbol \theta=(\alpha, \lambda_1, \beta)$, with the corresponding log-likelihood function 
\begin{align}
\log L(\boldsymbol \theta | Data) = & \, m (\log \alpha + \log \lambda_1) + m_2 \log \beta + (\alpha-1)\sum_{i=1}^{n}\log x_{i:m:n}\nonumber\\
& - \lambda_1(1+\beta)\sum_{i=1}^m (R_i^*+1)x_{i:m:n}^{\alpha}. \label{order-log-lik}
\end{align}
From (\ref{order-log-lik}), we can obtain the likelihood equation for $\lambda_1$, and by solving it for fixed $\alpha$ and $\beta$, we get the maximum likelihood estimate (MLE) for $\lambda$ as
\begin{equation}
\widehat{\lambda}_1(\alpha, \beta) = \frac{m}{(1+\beta)\sum_{i=1}^m(R_i^*+1)x_{i:m:n}^{\alpha}}. \label{order-lambda}
\end{equation}
Substituting $\widehat{\lambda}_1(\alpha, \beta)$ in (\ref{order-log-lik}), the
profile-log-likelihood function of $\alpha$ and $\beta$ is obtained as
\begin{equation}
p(\alpha, \beta) = p_1(\alpha) + p_2(\beta), \nonumber
\end{equation}
where 
\begin{align}
   p_1(\alpha) &= m \log \alpha - m \log(\sum_{i=1}^m(R_i^*+1)x_{i:m:n}^{\alpha}) +
   (\alpha-1)\sum_{i=1}^m\log x_{i:m:n},\label{eq:orderlikealpha}\\
   p_2(\beta) &= -m \log (1+\beta) + m_2 \log \beta.\nonumber
\end{align}
If $m_2<m_1$, the MLE of $\beta$ can be found by solving $\frac{\partial
p_2(\beta)}{\partial \beta}=0$. In this case, we have the MLE for $\beta$ as
\begin{equation*}
\widehat{\beta} = \frac{m_2}{m_1}.
\end{equation*}
Note that $p_2(\beta)$ is an increasing function of $\beta$ for $m_2\ge m_1$. Thus, the
MLE of $\beta$ in this case is 1. Clearly, the MLE for $\lambda_2$ can be obtained as 
\begin{equation*}
\widehat{\lambda}_2 = \widehat{\beta}\widehat{\lambda}_1.
\end{equation*}

\noindent {\sc Lemma}: $p_1(\alpha)$ is a unimodal function in $\alpha$. 

\noindent {\it Proof}: Note that
$$
p_1^\prime(\alpha) = \frac{m}{\alpha} - m\frac{\sum_{i=1}^m(R_i^*+1)x_{i:m:n}^{\alpha}\log x_{i:m:n}}{\sum_{i=1}^m(R_i^*+1)x_{i:m:n}^{\alpha}} + \sum_{i=1}^{m}\log x_{i:m:n},
$$ 
and 
\begin{equation}
p_1^{\prime\prime}(\alpha) = -\frac{m}{\alpha^2} - m\frac{g(\alpha)g''(\alpha) - \{g'(\alpha)\}^2}{\{g(\alpha)\}^2}, \nonumber
\end{equation}
where
\begin{equation}
g(\alpha) = \sum_{i=1}^m(R_i^*+1)x_{i:m:n}^{\alpha}, \quad g'(\alpha) = \sum_{i=1}^m(R_i^*+1)x_{i:m:n}^{\alpha}\log x_{i:m:n} \nonumber
\end{equation}
and
\begin{equation}
g''(\alpha) = \sum_{i=1}^m(R_i^*+1)x_{i:m:n}^{\alpha}(\log x_{i:m:n})^2. \nonumber
\end{equation}
Note that 
\begin{eqnarray}
& g(\alpha)g''(\alpha) - \{g'(\alpha)\}^2 &= \sum_{1 \leq i < j \leq
m}(R_i^*+1)(R_j^*+1)(\log x_{i:m:n}-\log x_{j:m:n})^2 \geq 0. \nonumber
\end{eqnarray}
Thus, $p_1(\alpha)$ is concave. Then, noting that $p_1(\alpha) \to -\infty$ as $\alpha \to
0$ or $\alpha \to \infty$, it follows immediately that $p_1(\alpha)$ is unimodal. \qed

Now, since $p_1(\alpha)$ is unimodal, to obtain MLE of $\alpha$, a simple one-dimensional
optimization technique like the Newton-Raphson, or the bisection method can be employed.
Alternatively, one can use a fixed-point equation approach like the following. Note that
equating $p^\prime(\alpha)$ to zero and rearranging the resulting equation, we have
\begin{equation*}
\alpha = \bigg[\frac{\sum_{i=1}^m(R_i^*+1)x_{i:m:n}^{\alpha}\log x_{i:m:n}}{\sum_{i=1}^m(R_i^*+1)x_{i:m:n}^{\alpha}} - \frac{1}{m}\sum_{i=1}^{m}\log x_{i:m:n}\bigg]^{-1} = h(\alpha).
\end{equation*}
Therefore, the following simple iterative algorithm is proposed to obtain MLEs of model
parameters.

\noindent {\sc Algorithm}: \newline 
\noindent {\sc Step 1}: Start with an initial value $\alpha^{(0)}$ \newline
\noindent {\sc Step 2}: Update by $\alpha^{(1)} = h(\alpha^{(0)})$ \newline
\noindent {\sc Step 3}: At the $(k+1)-$th step, obtain $\alpha^{(k+1)} = h(\alpha^{(k)})$ \newline
\noindent {\sc Step 4}: Stop when $|\alpha^{(k+1)} - \alpha^{(k)}| < \epsilon$, for some pre-fixed $\epsilon > 0$, and take $\widehat{\alpha} = \alpha^{(k+1)}$ \newline
\noindent {\sc Step 5}: Calculate $\widehat{\beta} = \begin{cases}
   \frac{m_2}{m_1} & \text{if }m_1>m_2\\ 1 & \text{if } m_1\le m_2
\end{cases}$ \newline
\noindent {\sc Step 6}: From (\ref{order-lambda}), calculate
$\widehat{\lambda}_1(\widehat{\alpha},\widehat{\beta})$ \newline
\noindent {\sc Step 7}: Finally, calculate $\widehat{\lambda}_2 =
\widehat{\beta}\widehat{\lambda}_1$.

\subsection{\sc Bayesian inference}\label{subsec:BIOrder}

Note that the MLEs of the unknown parameters do not exist in closed from, and hence the
further analyses are based on the asymptotic properties of the MLEs. Therefore, it seems
that the Bayesian inference is a natural alternative. In this subsection, we consider
Bayesian inference of adaptive progressively Type-II censored competing risks data under
the order restriction $\lambda_1>\lambda_2$, with the same reparameterization
$\lambda_2=\beta\lambda_1$, $0<\beta\le 1$. Following the approach of Berger and
Sun~\cite{BS1993}, Kundu and Gupta~\cite{KG2006}, and Kundu~\cite{K2008}, here it is
assumed that $\lambda_1$ has a gamma prior with shape $a_1>0$ and scale $b_1>0$. Thus, the
prior probability density function (PDF) of $\lambda_1$ is given by
\begin{align*}
\pi_1(\lambda_1)\propto\lambda_1^{a_1-1}e^{-b_1\lambda_1} \quad \text{ for } \quad \lambda_1>0.
\end{align*}
We also assume that the prior distribution of the shape parameter $\alpha$ is a gamma
distribution with shape $a_2>0$ and scale $b_2>0$ having PDF
\begin{align*}
\pi_2(\alpha)\propto\alpha^{a_2-1}e^{-b_2\alpha} \quad \text{ for } \quad \alpha>0. 
\end{align*}
As $\beta\in(0,\,1]$ and beta distribution is a quite flexible distribution with support
(0, 1], we assume that $\beta$ has a beta prior with hyper parameters $a_3>0$ and $b_3>0$
and PDF
\begin{align*}
\pi_3(\beta)\propto\beta^{a_3-1}(1-\beta)^{b_3-1} \quad \text{ for } \quad 0<\beta\le 1.
\end{align*}
It is also assumed that $\lambda_1$, $\alpha$, and $\beta$ are apriori independently
distributed.  Now, based on the likelihood function in \eqref{order-lik}, the priors
$\pi_1(\cdot)$, $\pi_2(\cdot)$, and $\pi_3(\cdot)$, the posterior PDF of $\lambda_1$,
$\alpha$, and $\beta$ can be written as
\begin{align}
\widetilde\pi_1(\lambda_1,\,\alpha,\,\beta\vert Data) \propto \lambda_1^{m+a_1-1}
\alpha^{m+a_2-1} \beta^{m_2+a_3-1} (1-\beta)^{b_3-1}
e^{-A_1(\alpha,\,\beta)\lambda_1-A_2\alpha}, \label{eq:PostOrder}
\end{align}
for $\lambda_1>0$, $\alpha>0$, and $0<\beta<1$, where
$A_1(\alpha,\,\beta)=b_1+(1+\beta)\sum_{i=1}^m(R^*_i+1)x_{i:m:n}^\alpha$ and $A_2=b_2 -
\sum_{i=1}^{m} \log x_{i:m:n}$. The Bayes estimate (BE) of some parametric function
$h(\lambda_1,\,\alpha,\,\beta)$ under squared error loss function is given by
\begin{align}
\widehat h_{BE}(\lambda_1,\,\alpha,\,\beta) = \int_0^\infty\int_0^\infty\int_0^1
h(\lambda_1,\,\alpha,\,\beta)\widetilde\pi_1(\lambda_1,\,\alpha,\,\beta\vert
Data)d\beta d\lambda_1 d\alpha.\label{eq:BEOrder}
\end{align}
In general the integration in \eqref{eq:BEOrder} does not exist in close form. Hence, we
propose a simulation consistent algorithm based on importance sampling technique to
compute BE and to construct credible interval (CRI) of some parametric function. 

Let $T_1,\,\ldots,\,T_l$ denote a random sample of size $l$ from a We$(\nu,\,\lambda)$
distribution having CDF $F(\cdot)$. As one can write
$$\ln\left(-\log\left(1-F(x)\right)\right)=\ln\lambda+\nu\ln x,$$
$\nu$ can be estimated using a simple linear regression $$y_i=\mu+\nu z_i+e_i,$$ where
$y_i=\ln\left(-\log\left(1-\frac{i-0.5}{l}\right)\right)$, $z_i=\ln T_{i:l}$, and
$\mu=\ln\lambda$. We use this method to find an approximate estimate for $\alpha$, which
will be used to generate sample using importance sampling scheme. Let $\widetilde\alpha_1$
and $\widetilde\alpha_2$ be the estimates of $\alpha$ that are found using the linear
regression method based on the failures corresponding to $I_1$ and $I_2$, respectively.
Also define
$\widetilde\alpha=\frac{1}{2}\left(\widetilde\alpha_1+\widetilde\alpha_2\right)$.

Note that the posterior PDF given in \eqref{eq:PostOrder} can be expressed as
\begin{align*}
{} & \widetilde\pi_1(\lambda_1,\,\alpha,\,\beta\vert Data)\propto
w(\alpha,\,\beta)\widetilde\pi_4(\lambda_1\vert\alpha,\,\beta) \widetilde\pi_3(\beta)
\widetilde\pi_2(\alpha),
\intertext{where}
{} & w(\alpha,\,\beta) = \alpha^{m+a_2-2} \beta^{m_2+a_3-1}(1-\beta)^{b_3-1}
e^{-\alpha(A_2-b)} A_1^{-(m+a_1)}(\alpha,\,\beta),\\
{} & \widetilde\pi_2(\alpha) = b^2 \alpha e^{-b\alpha},\\
{} & \widetilde\pi_3(\beta)=1,\\
{} & \widetilde\pi_4(\lambda_1\vert\alpha,\,\beta) =
\frac{A_1^{m+a_1}(\alpha,\,\beta)}{\Gamma(m+a_1)} \lambda_1^{m+a_1-1}
e^{-A_1(\alpha,\,\beta)\lambda_1},\\
{} & b = \frac{2}{\widetilde\alpha}.
\end{align*}

Therefore, the following algorithm is proposed to obtain BE and CRI of a parametric
function, say $h\left( \lambda_1,\,\alpha,\,\beta \right)$.

\noindent {\sc Algorithm}:\newline
\noindent {\sc Step 1}: Generate $\alpha_1$ from $\widetilde\pi_2(\cdot)$.\newline
\noindent {\sc Step 2}: Generate $\beta_1$ from $\widetilde\pi_3(\cdot)$.\newline
\noindent {\sc Step 3}: Generate $\lambda_{11}$ from $\widetilde\pi_4(\cdot\vert\alpha_1,\,\beta_1)$.\newline
\noindent {\sc Step 4}: Repeat the steps 1, 2, and 3, $M$ times to get $(\lambda_{1i},\,\alpha_i,\,\beta_i)$, $i=1,\,2,\,\ldots,\,M$. \newline
\noindent {\sc Step 5}: Calculate $w_{i}=w(\alpha_i,\,\beta_i)$ for $i=1,\,2,\,\ldots,\,M$. \newline
\noindent {\sc Step 6}: Calculate $w^*_{i}=\frac{w_{i}}{\sum_{j=1}^M w_{j}}$ for $i=1,\,2,\,\ldots,\,M$. \newline
\noindent {\sc Step 7}: Calculate $h_i=h(\lambda_{1i},\,\alpha_i,\,\beta_i)$ for $i=1,\,2,\,\ldots,\,M$. \newline
\noindent {\sc Step 8}: Approximate $\widehat h_{BE}(\lambda_1,\,\alpha,\,\beta)$ by
$\sum_{i=1}^M w^*_{i}h_i$.\newline
\noindent {\sc Step 9}: Order $h_i$'s in ascending order to obtain $h_{(1)}\leq
h_{(2)}\leq \ldots<h_{(M)}$. Order $w^*_{i}$'s accordingly to get
$w_{(1)},\,w_{(2)},\,\ldots,\,w_{(M)}$. Note that $w_{(1)}$, $w_{(2)}$, $\ldots$,
$w_{(M)}$ may not be ordered. \newline
\noindent {\sc Step 10}: Construct a $100(1-\gamma)\%$ CRI as $[h_{(j_1)},\,h_{(j_2)}]$, where $j_1<j_2$ satisfy
\begin{align}
\sum_{i=j_1}^{j_2}w_{(i)}\leq1-\gamma<\sum_{i=j_1}^{j_2+1}w_{(i)}.\label{eq:CRIOrder}
\end{align}
\noindent {\sc Step 11}: Construct the $100(1-\gamma)\%$ height posterior density CRI as
$[h_{(j_1^*)},\,h_{(j_2^*)}]$, where $j_1^*<j_2^*$ satisfy
\begin{align*}
\sum_{i=j_1^*}^{j_2^*}w_{(i)}\leq1-\gamma<\sum_{i=j_1^*}^{j_2^*+1}w_{(i)}\text{  and  } h_{(j_2^*)}-h_{(j_1^*)}\leq h_{(j_2)}-h_{(j_1)},
\end{align*}
for all $j_1$ and $j_2$ satisfying \eqref{eq:CRIOrder}.

Note that choices of $\widetilde\pi_2(\alpha),\, \widetilde\pi_3(\beta),\,
\pi_4(\lambda_1\vert\alpha,\,\beta)$, and $w(\alpha,\,\beta)$ may not be optimal, but we
have noticed that these choices work quite well.  The trivial choice of
$\widetilde\pi_2(\cdot)$ would be a gamma PDF with shape $m+a_2-1$ and scale $A_2$.
However, this choice needs the re-scaling of original data points so that scale parameter
$A_2>0$. We have also noticed that with the trivial choice of $\widetilde\pi_2(\cdot)$ the
generated values of $\alpha$, in some cases, are such that the weight $w(\alpha,\,\beta)$
concentrates on one or two points. Hence, we find a crude estimate of $\alpha$ using the
liner regression technique as described above and choose $\widetilde\pi_2(\cdot)$ such
that the mean of the PDF is $\widetilde\alpha$.

\section{\sc Inference without order restriction}\label{sec:InfNoOrder}

Ren and Gui~\cite{RG2021} considered inferential issues under a similar setup. The authors
assumed that the latent lifetimes follow Weibull distributions with different shape and
scale parameters under different risk factors. It may be noted that in this case the
profile log-likelihood function of the shape parameters can be expressed as a sum of two
functions, where the first function is the profile log-likelihood function of shape
parameter corresponding to the first latent failure time, and the second function is the
profile log-likelihood function of shape parameter corresponding to the second latent
failure time. Consequently, the implementation of inferential techniques becomes easier
when shape parameters are assumed to be different compared to when shape parameters are
assumed to be same.  As we will compare the order restricted inference with unrestricted
inference, in this section we briefly describe the inferential techniques when there is no
order restriction on parameters of the lifetime distributions of the competing causes and
when shape parameters are assumed to be same.  Here also, all derivations are carried out
under Case-II of adaptive progressive censoring as it includes Case-I as a special case. 

\subsection{\sc Likelihood inference}\label{subsec:LikInfNoOrder}

In this case, the log-likelihood function of $\boldsymbol{\zeta} = (\alpha,\, \lambda_1,\,
\lambda_2)$ is
\begin{align}
\log L(\boldsymbol\zeta | Data) = &\,m \log \alpha + m_1 \log \lambda_1 + m_2 \log \lambda_2 + (\alpha-1)\sum_{i=1}^{m}\log x_{i:m:n}\nonumber\\
 & - (\lambda_1+\lambda_2)\sum_{i=1}^m (R_i^*+1)x_{i:m:n}^{\alpha}. \label{log-lik}
\end{align}
For fixed $\alpha$, equating the first-order derivatives of the log-likelihood function in
(\ref{log-lik}) with respect to $\lambda_1$ and $\lambda_2$ to zero, we obtain
\begin{equation*}
\widehat{\lambda}_1(\alpha) = \frac{m_1}{\sum_{i=1}^m(R_i^*+1)x_{i:m:n}^{\alpha}} \quad
\text{and} \quad \widehat{\lambda}_2(\alpha) =
\frac{m_2}{\sum_{i=1}^m(R_i^*+1)x_{i:m:n}^{\alpha}}. \label{lambdas}
\end{equation*}
Substituting $\widehat{\lambda}_1(\alpha)$ and $\widehat{\lambda}_2(\alpha)$ in
(\ref{log-lik}), the profile log-likelihood in $\alpha$ is obtained as
\begin{equation*}
p_3(\alpha) = m \log \alpha - m \log\bigg(\sum_{i=1}^m(R_i^*+1)x_{i:m:n}^{\alpha}\bigg) +
(\alpha-1)\sum_{i=1}^{m}\log x_{i:m:n}, \label{profile-alpha}
\end{equation*}
which is same as the profile log-likelihood function $p_1(\alpha)$ as given in
\eqref{eq:orderlikealpha}. Therefore, $p_3(\alpha)$ is a unimodal function in $\alpha$,
and hence, the MLE of $\alpha$ can easily be obtained using a one-dimensional optimization
technique.  Once the MLE of $\alpha$ is obtained, the MLEs of $\lambda_1$ and $\lambda_2$
can be obtained as $\widehat{\lambda}_1\left( \widehat{\alpha} \right)$ and
$\widehat{\lambda}_2\left( \widehat{\alpha} \right)$, respectively, where
$\widehat{\alpha}$ is the MLE of $\alpha$.  


\subsection{\sc Bayesian inference}\label{sec:BayesUnorder}

Here it is assumed that $\lambda_1$, $\lambda_2$, and $\alpha$ have  gamma priors with the
prior PDFs
\begin{align*}
\pi_4(\lambda_1)\propto\lambda_1^{a_4-1}e^{-b_4\lambda_1} \text{ for } \lambda_1>0,\\
\pi_5(\lambda_2)\propto\lambda_2^{a_5-1}e^{-b_5\lambda_2} \text{ for } \lambda_2>0,\\
\pi_6(\alpha)\propto \alpha^{a_6-1}e^{-b_6\alpha} \text{ for } \alpha>0.
\end{align*}
It is further assumed that $\alpha$, $\lambda_1$, and $\lambda_2$ are independently
distributed. Now, the joint posterior PDF of $\alpha$, $\lambda_1$, and $\lambda_2$ can be
expressed as follows: For $\lambda_1>0$, $\lambda_2>0$, and $\alpha>0$,
\begin{align*}
&\widetilde\pi_5(\lambda_1,\,\lambda_2,\,\alpha\vert
Data)\propto\lambda_1^{m_1+a_4-1}\lambda_2^{m_2+a_5-1}\alpha^{m+a_6-1}
e^{-A_3(\alpha)\lambda_1-A_4(\alpha)\lambda_2-A_5\alpha}, \label{eq:PosUnorder}
\end{align*}
where $A_3(\alpha)=b_4+\sum_{i=1}^m(R_i^*+1)x_{i:m:n}^\alpha$,
$A_4(\alpha)=b_5+\sum_{i=1}^m(R_i^*+1)x_{i:m:n}^\alpha$, and $A_5=b_6-\sum_{i=1}^m\log
x_{i:m:n}$. The BE of some function of $\lambda_1$,
$\lambda_2$, and $\alpha$, say $g(\lambda_1,\,\lambda_2,\,\alpha)$, under squared error
loss function is the posterior expectation of $g(\lambda_1,\,\lambda_2,\,\alpha)$, which
is given by
\begin{align*}
\widehat g_{BE}(\lambda_1,\,\lambda_2,\,\alpha) = \int_0^\infty\int_0^\infty\int_0^\infty
g(\lambda_1,\,\lambda_2,\,\alpha) \widetilde\pi_5(\lambda_1,\,\lambda_2,\,\alpha\vert
Data)d\lambda_1 d\lambda_2 d\alpha,
\end{align*}
provided it exists. Note that thee posterior PDF of $\lambda_1,\,\lambda_2$, and $\alpha$
can be rewritten as follows:
\begin{align*}
\widetilde\pi_5(\lambda_1,\,\lambda_2,\,\alpha\vert Data)\propto
v(\alpha) \widetilde\pi_6(\lambda_1\vert\alpha) \widetilde\pi_7(\lambda_2\vert\alpha)
\widetilde\pi_2(\alpha),
\end{align*}
where
\begin{align*}
& v(\alpha)=\alpha^{m+a_6-2}e^{-\alpha(A_5-b)} A_3^{-(m_1+a_4)}(\alpha) A_4^{-(m_2+a_5)}(\alpha),\\
& \widetilde\pi_6(\lambda_1\vert\alpha) = \frac{A_3^{m_1+a_4}(\alpha)}{\Gamma(m_1+a_4)}
\lambda_1^{m_1+a_4-1} e^{-\lambda_1 A_3(\alpha)},\\
& \widetilde\pi_7(\lambda_2\vert\alpha) = \frac{A_4^{m_2+a_5}(\alpha)}{\Gamma(m_2+a_5)}
\lambda_2^{m_2+a_5-1} e^{-\lambda_2 A_4(\alpha)},
\end{align*}
and $\widetilde\pi_2(\cdot)$ is defined in Section~\ref{subsec:BIOrder}. Now, a simulation
consistent algorithm, like the previous section, can be used to compute BEs and to
construct CRIs of the model parameters.

\section{\sc Simulation study}\label{sec:Simu}

Computational works for this article have been carried out by using the R software. For
each unit, two lifetimes corresponding to the two independent competing causes of failure
are generated from Weibull distributions with different scale parameters and same shape
parameter. The lifetime and cause of failure of a unit are then determined by identifying
the minimum of the two lifetimes. The total number of units on test, i.e., $n$ is taken as
50, and the number of observed failures, i.e., $m$ is taken as 40. Progressive Type-II
censoring is incorporated into the simulation study according to three different schemes,
namely, $\boldsymbol R=$ $(0,...,0,10)$, $(10,0,...,0)$, and $(0,...,10,...,0)$, i.e., the
right censoring, first step censoring plan (FSP), and one step censoring plan (OSP),
respectively. Two different values for the time controlling parameter $T$ are considered,
namely, 0.25 and 0.75. Two values for the shape parameter have been chosen, namely, 0.5
and 1.5, as they correspond to two very different shapes for the Weibull distributions.
For each value of the shape parameter, the scale parameters are taken as $\left(
\lambda_1,\, \lambda_2 \right) = (1.2,\,1)$ and $(1.4,\,1)$. Note that the values of
$\beta$ are $\frac{1}{1.2}$ and $\frac{1}{1.4}$ for $(\lambda_1,\,\lambda_2) = \left(
1.2,\, 1 \right)$ and $\left( 1.4,\,1 \right)$, respectively. 

Parametric bootstrap confidence intervals for several parameters are computed and
performances are judged through extensive numerical simulation. To construct parametric
bootstrap confidence interval of a parametric function, say $\tau(\cdot)$, $B$ bootstrap
MLEs of $\tau(\cdot)$ are calculated. Let these MLEs be denoted by $\widehat{\tau}^*_1,\,
\widehat{\tau}^*_2,\, \ldots,\, \widehat{\tau}^*_B$. Percentile bootstrap confidence
intervals are obtained simply by choosing appropriate percentiles of bootstrap MLEs.
Alternatively, one may consider the following bootstrap confidence interval, which will be
called parametric bootstrap confidence interval in this article to distinguish from
percentile bootstrap confidence interval. A $100\left( 1-\gamma \right)\%$ parametric
bootstrap confidence interval is given by
$ (\widehat{\tau} - b_{\tau} - z_{\gamma/2}\sqrt{v_{\tau}},\, \widehat{\tau} - b_{\tau} +
z_{\gamma/2}\sqrt{v_{\tau}})$, 
where $b_{\tau} = \overline{\widehat{\tau}^*}-\widehat{\tau}, \quad v_{\tau} =
\frac{1}{B-1}\sum_{i=1}^{B}\left(\widehat{\tau}_{i}^* -
\overline{\widehat{\tau}^*}\right)^2, \text{ and } \overline{\widehat{\tau}^*} =
\frac{1}{B}\sum_{i=1}^{B}\widehat{\tau}_{i}^*$.

Tables~\ref{tab:order1} -- \ref{tab:order4} show performances of point and interval
estimates corresponding to likelihood and Bayesian inference under the order restricted
setup. In these tables, we report bias and means square error (MSE) for both MLEs and BEs
of the relevant parameters. The nominal level for bootstrap confidence intervals and CRIs
is taken to be 95\%. The coverage probabilities and average lengths of different intervals
are reported. The coverage probabilities are abbreviated as CPB for parametric bootstrap
confidence interval, CPP for percentile bootstrap confidence interval, CPS for symmetric
CRI, and CPH for highest posterior density CRI. Similarly, average lengths are
abbreviated. To compare the performance of order restricted inference with unrestricted
inference, we also compute same set of performance measures when no order restriction are
imposed and the corresponding results are reported in Tables~\ref{tab:unorder1} --
\ref{tab:unorder4}.

In Tables~\ref{tab:order1} -- \ref{tab:order4}, we notice that performance of the MLEs and
the Bayes estimates of the scale parameters are quite comparable with respect to bias and
mean squared error (MSE), while the Bayes estimate for the shape parameter $\alpha$ is
better than the corresponding MLE, specially for OSP.  In these tables, we notice that
coverage probabilities and average lengths of the two Bayesian credible intervals are
quite similar, and the coverage probabilities for these intervals are very close to the
nominal confidence level 95\%. However, the two bootstrap confidence intervals are
somewhat different than the other intervals. The parametric bootstrap, and the
percentile bootstrap confidence intervals seem to be wider than the CRIs. Particularly for
the shape parameter $\alpha$, the average length of the parametric bootstrap
confidence interval is significantly larger than that for the other intervals for
$\alpha$. As a result, the coverage probability of the parametric bootstrap confidence
interval for $\alpha$ is significantly above the nominal confidence level. Moreover, this
phenomenon is particularly true for FSP and OSP.  Different magnitudes of the test time
controlling parameter $T$ do not seem to have any impact on the coverage probabilities and
average lengths of different confidence intervals.               

The simulation results for the unrestricted case are presented in Tables
\ref{tab:unorder1} --\ref{tab:unorder4}. By comparing the entries of
Tables~\ref{tab:order1} and \ref{tab:unorder1}, it is observed that the performances of
MLE and BE of $\lambda_2$ (scale parameter corresponding to non-dominating risk factor)
improve significantly in the presence of order restriction on the scale parameters. The
performances of MLE and BE of $\lambda_1$ (scale parameter corresponding to dominating
risk factor) also improve to some extend when order restriction is imposed. The
performances of the estimators of shape parameter improve significantly under the order
restriction for FSP, while for Type-II censoring or for OSP, the performance are seems to
be equivalent. The same trend is noticed for other choices of the parameters.


\section{\sc Illustrative example}\label{sec:Illus}

In this section, we present analysis of a dataset to numerically illustrate the methods of
inference developed in this paper. The dataset is simulated using the procedure described
in Section~\ref{sec:Simu}. For this dataset, the total number of units on test, i.e., $n$,
is 100 while the number of observed failures, i.e., $m$, is 90. The two Weibull
distributions corresponding to the two competing risk factors are taken as We$(\alpha=1.5,
\lambda_1=1.5)$, and We$(\alpha=1.5,\lambda_2=1.0)$. The one-step censoring plan (OSP),
i.e., $\boldsymbol R = (0,...,10,...,0)$ is used to incorporate progressive Type-II
censoring. The time controlling parameter $T$ is taken as 0.5. 

It is observed that the number of failures from the two causes, i.e., $m_1$ and $m_2$, are
58 and 32, respectively. The mean lifetimes corresponding to the two causes of failure are
0.464 and 0.459, respectively, with respective standard deviations 0.343 and 0.290. 

Given the data, it may be of interest to know whether the two scale parameters for the two
competing risk factors are really different or not. If the two scale parameters cannot be
taken to be different, then the competing risks modelling is not meaningful in this case.
We thus test a hypothesis 
\begin{equation}
H_o: \lambda_1 = \lambda_2 = \lambda (\textrm{say}), \quad \textrm{against} \quad H_1: \lambda_1 \neq \lambda_2. \nonumber
\end{equation}
This hypotheses can be tested using likelihood ratio test (LRT). The test statistic is computed as 
\[
\Lambda = -2(\hat{l}_{H_o} - \hat{l}_{H_o\cup H_1}) = 7.619
\] 
where $\hat{l}_{H_o}$ and $\hat{l}_{H_o\cup H_1}$ are the maximized log-likelihood values
for the restricted and unrestricted models, respectively. It is known that $\Lambda$
follows asymptotically a $\chi^2$-distribution with one degree of freedom under null
hypothesis. At 5\% level this null hypothesis of equality is rejected, as $\chi^2_{0.05;1}
= 3.84$. 

\begin{figure}[ht]
   \centering
   \includegraphics{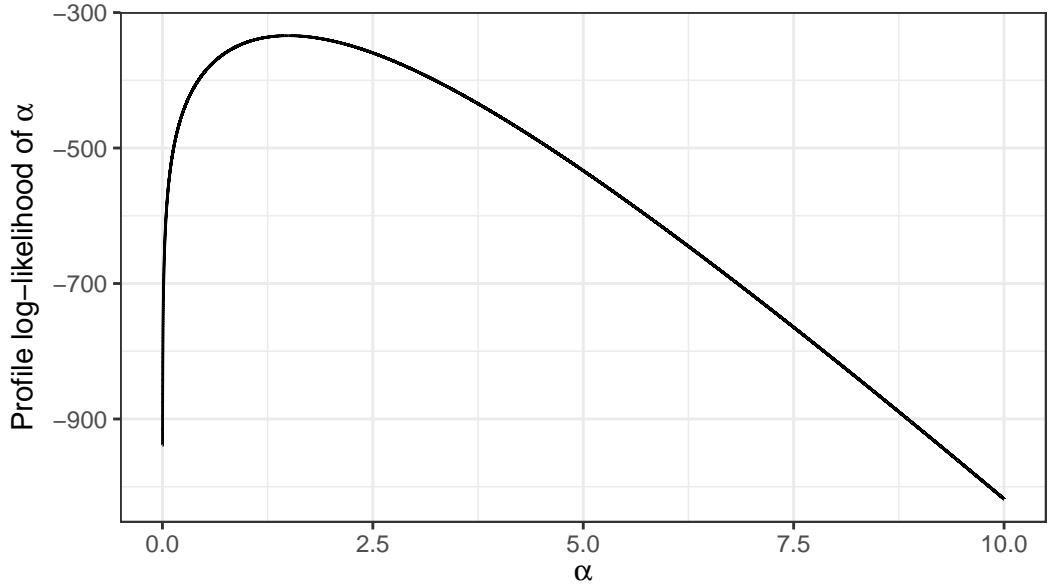}
   \caption{Plot of profile log-likelihood function of $\alpha$ under order restriction}
   \label{fig:DataAnaProf}
\end{figure}

The plot of profile log-likelihood function, $p_1(\alpha)$, is provided in the
Figure~\ref{fig:DataAnaProf}. It is clear from the figure that the MLE of $\alpha$ is near
to 1.5. We take 1.5 as initial guess to implement iterative method to find the MLE of
$\alpha$ by maximizing \eqref{eq:orderlikealpha}. The point and interval estimates of the
model parameter based on this dataset with order restriction on the scale parameters are
reported in Table~\ref{tab:DataAnalysisOrder}. Table~\ref{tab:DataAnalysisUnorder} gives
the point and interval estimates for the model parameters obtained based on this dataset
when no restriction is imposed on the scale parameters.


\section{\sc Conclusion}\label{sec:Con}

In this article, we have discussed analysis of adaptive Type-II progressively censored
competing risks data under order restriction on the scale parameters. The order
restriction comes naturally if it is known that one risk factor dominates the other. The
MLEs of the model parameters are derived, and parametric bootstrap confidence intervals
are obtained. Bayes estimates of the model parameters are also derived, and construction
of symmetric and highest posterior density credible intervals are discussed. Through an
extensive Monte Carlo simulation study, the performance of the proposed methods are
assessed. We also compare the performances of different estimators of different parameters
under order restriction with those in unrestricted situation.

Through the simulation study, we observe that the Bayesian methods perform better than
the classical methods at least for small samples under order restriction on the scale
parameters. It is also noticed that the performances of estimators of scale parameter
corresponding to lifetime of the non-dominating risk factor improve significantly when
order restriction is used. The performances of the estimators of scale parameter
corresponding to lifetime of the dominating risk factor are also improved. The
performances of the estimators of the common shape parameter improve significantly under
the order restriction for FSP, while for Type-II censoring or for OSP, the performances
seem to be equivalent.

If we assume that the latent failure times have $We(\alpha_1,\,\lambda_1)$ and $We\left(
\alpha_2,\,\lambda_2 \right)$ distributions, respectively, under risk factors 1 and 2, it
is quite difficult to find a necessary and sufficient condition such that the mean lifetime
under risk factor 1  is less than that under risk factor 2. However, one may consider a
sufficient condition on the parameters such that ordering on mean lifetimes holds true. One
such sufficient condition can described as follows. Let us reparameterize as
$\theta_i=\lambda_i^{\frac{1}{\alpha_i}}$ for $i=1,\,2$. Then, $\alpha_1>\alpha_2$ and
$\theta_1>\theta_2$ is a set of sufficient conditions. Under this conditions, the analysis
can be performed in a similar way as described in Section~\ref{subsec:BIOrder}. More work
is needed along this direction.

\begin{table}\tiny
   \caption{Performance of MLEs, Bayes estimates, confidence intervals, and credible
   intervals for $(\alpha, \lambda_1, \beta)$ = (1.5, 1.2, 1/1.2) under order restriction.}
   \label{tab:order1}
   \begin{center}
      \begin{tabular}{*{15}{c}}
         \toprule
         \multicolumn{3}{c}{} & \multicolumn{6}{c}{Likelihood Based} & \multicolumn{6}{c}{Bayesian}\\
         \cmidrule(lr){4-9}\cmidrule(lr){10-15}
         $\boldsymbol R$ & $T$ & Par & Bias & MSE & CPB & ALB & CPP & ALP & Bias & MSE & CPS & ALS & CPH & ALH \\ 
         \midrule
         \multirow{8}{*}{(0,...,0,10)}         & \multirow{4}{*}{0.25} & $\alpha$    &  0.07 & 0.055 & 0.956 & 0.91 & 0.920 & 0.90 &  0.05 & 0.052 & 0.944 & 0.83 & 0.947 & 0.82\\
                                               &                       & $\lambda_1$ &  0.12 & 0.128 & 0.986 & 1.53 & 0.884 & 1.48 &  0.17 & 0.131 & 0.950 & 1.15 & 0.962 & 1.12\\
                                               &                       & $\lambda_2$ &  0.04 & 0.081 & 0.934 & 1.13 & 0.971 & 1.12 & -0.01 & 0.060 & 0.953 & 0.91 & 0.942 & 0.88\\
                                               &                       & $\beta$     & -0.03 & 0.034 & 0.896 & 0.71 & 0.939 & 0.63 & -0.09 & 0.018 & 0.985 & 0.52 & 0.968 & 0.49\\\cmidrule(lr){2-15}
                                               & \multirow{4}{*}{0.75} & $\alpha$    &  0.06 & 0.051 & 0.955 & 0.90 & 0.921 & 0.89 &  0.06 & 0.051 & 0.950 & 0.83 & 0.950 & 0.82\\
                                               &                       & $\lambda_1$ &  0.11 & 0.114 & 0.995 & 1.51 & 0.882 & 1.46 &  0.17 & 0.134 & 0.944 & 1.16 & 0.960 & 1.12\\
                                               &                       & $\lambda_2$ &  0.05 & 0.083 & 0.936 & 1.16 & 0.960 & 1.14 & -0.01 & 0.058 & 0.946 & 0.91 & 0.939 & 0.88\\
                                               &                       & $\beta$     & -0.02 & 0.035 & 0.876 & 0.71 & 0.922 & 0.62 & -0.09 & 0.018 & 0.982 & 0.52 & 0.967 & 0.49\\\midrule
         \multirow{8}{*}{(10,0,...,0)}         & \multirow{4}{*}{0.25} & $\alpha$    &  0.12 & 0.362 & 0.983 & 2.72 & 0.915 & 1.27 &  0.04 & 0.041 & 0.937 & 0.73 & 0.937 & 0.72\\
                                               &                       & $\lambda_1$ &  0.08 & 0.093 & 0.982 & 1.46 & 0.915 & 1.27 &  0.13 & 0.089 & 0.942 & 0.98 & 0.951 & 0.95\\
                                               &                       & $\lambda_2$ & -0.00 & 0.062 & 0.944 & 1.05 & 0.957 & 1.01 & -0.04 & 0.043 & 0.928 & 0.78 & 0.918 & 0.76\\
                                               &                       & $\beta$     & -0.03 & 0.032 & 0.910 & 0.71 & 0.946 & 0.62 & -0.09 & 0.018 & 0.982 & 0.52 & 0.966 & 0.49\\\cmidrule(lr){2-15}
                                               & \multirow{4}{*}{0.75} & $\alpha$    &  0.12 & 0.394 & 0.992 & 2.80 & 0.908 & 1.31 &  0.04 & 0.038 & 0.948 & 0.73 & 0.943 & 0.72\\
                                               &                       & $\lambda_1$ &  0.06 & 0.093 & 0.979 & 1.35 & 0.911 & 1.27 &  0.13 & 0.088 & 0.942 & 0.99 & 0.948 & 0.95\\
                                               &                       & $\lambda_2$ &  0.00 & 0.057 & 0.943 & 1.04 & 0.961 & 1.01 & -0.04 & 0.042 & 0.935 & 0.78 & 0.924 & 0.76\\
                                               &                       & $\beta$     & -0.02 & 0.033 & 0.908 & 0.71 & 0.944 & 0.62 & -0.09 & 0.019 & 0.982 & 0.52 & 0.967 & 0.49\\\midrule
         \multirow{8}{*}{(0,...,10,...,0)} & \multirow{4}{*}{0.25}     & $\alpha$    &  0.05 & 0.051 & 0.976 & 0.91 & 0.931 & 0.89 &  0.06 & 0.053 & 0.944 & 0.83 & 0.946 & 0.82\\
                                           &                           & $\lambda_1$ &  0.12 & 0.113 & 0.991 & 1.52 & 0.894 & 1.46 &  0.17 & 0.125 & 0.944 & 1.15 & 0.963 & 1.12\\
                                           &                           & $\lambda_2$ &  0.03 & 0.071 & 0.947 & 1.14 & 0.974 & 1.12 & -0.01 & 0.055 & 0.954 & 0.91 & 0.948 & 0.88\\
                                           &                           & $\beta$     & -0.02 & 0.033 & 0.899 & 0.71 & 0.945 & 0.62 & -0.09 & 0.019 & 0.982 & 0.52 & 0.965 & 0.49\\\cmidrule(lr){2-15}
                                           & \multirow{4}{*}{0.75}     & $\alpha$    &  0.09 & 0.127 & 0.980 & 1.41 & 0.891 & 1.01 &  0.05 & 0.040 & 0.949 & 0.73 & 0.948 & 0.72\\
                                           &                           & $\lambda_1$ &  0.10 & 0.103 & 0.987 & 1.44 & 0.890 & 1.37 &  0.15 & 0.101 & 0.949 & 1.04 & 0.959 & 1.01\\
                                           &                           & $\lambda_2$ &  0.02 & 0.067 & 0.928 & 1.08 & 0.957 & 1.05 & -0.03 & 0.046 & 0.948 & 0.82 & 0.939 & 0.80\\
                                           &                           & $\beta$     & -0.03 & 0.033 & 0.908 & 0.71 & 0.943 & 0.62 & -0.09 & 0.019 & 0.980 & 0.52 & 0.967 & 0.49\\
                                           \bottomrule
      \end{tabular}
   \end{center}
\end{table}  

\begin{table}\tiny
    \caption{Performance of MLEs, Bayes estimates, confidence intervals, and credible
    intervals for $(\alpha, \lambda_1, \beta)$ = (0.5, 1.2, 1/1.2) under order
    restriction.}
   \label{tab:order2}
	\begin{center}
		\begin{tabular}{*{15}{c}}
			\toprule
            \multicolumn{3}{c}{} & \multicolumn{6}{c}{Likelihood Based} & \multicolumn{6}{c}{Bayesian}\\
            \cmidrule(lr){4-9}\cmidrule(lr){10-15}
         $\boldsymbol R$ & $T$ & Par & Bias & MSE & CPB & ALB & CPP & ALP & Bias & MSE & CPS & ALS & CPH & ALH \\ 
			\midrule
         \multirow{8}{*}{(0,...,0,10)}         & \multirow{4}{*}{0.25} & $\alpha$    &  0.02 & 0.006 & 0.966 & 0.30 & 0.916 & 0.30 &  0.02 & 0.006 & 0.947 & 0.28 & 0.947 & 0.27\\
			                                      &                       & $\lambda_1$ &  0.12 & 0.117 & 0.989 & 1.53 & 0.879 & 1.48 &  0.16 & 0.125 & 0.951 & 1.15 & 0.959 & 1.10\\
                                               &                       & $\lambda_2$ &  0.05 & 0.072 & 0.950 & 1.16 & 0.975 & 1.14 & -0.01 & 0.055 & 0.959 & 0.90 & 0.949 & 0.87\\
			                                      &                       & $\beta$     & -0.02 & 0.032 & 0.904 & 0.71 & 0.950 & 0.62 & -0.09 & 0.018 & 0.989 & 0.52 & 0.974 & 0.49\\\cmidrule(lr){2-15}
			                                      & \multirow{4}{*}{0.75} & $\alpha$    &  0.02 & 0.006 & 0.961 & 0.30 & 0.922 & 0.30 &  0.02 & 0.006 & 0.942 & 0.28 & 0.943 & 0.27\\
			                                      &                       & $\lambda_1$ &  0.12 & 0.126 & 0.991 & 1.54 & 0.867 & 1.49 &  0.16 & 0.124 & 0.950 & 1.15 & 0.963 & 1.11\\
                                               &                       & $\lambda_2$ &  0.03 & 0.070 & 0.939 & 1.15 & 0.961 & 1.13 & -0.01 & 0.055 & 0.952 & 0.90 & 0.943 & 0.87\\
			                                      &                       & $\beta$     & -0.03 & 0.034 & 0.887 & 0.71 & 0.936 & 0.62 & -0.09 & 0.019 & 0.984 & 0.52 & 0.970 & 0.49\\\midrule
			\multirow{8}{*}{(10,0,...,0)}         & \multirow{4}{*}{0.25} & $\alpha$    &  0.04 & 0.041 & 0.993 & 0.92 & 0.901 & 0.37 &  0.02 & 0.004 & 0.952 & 0.25 & 0.950 & 0.24\\
			                                      &                       & $\lambda_1$ &  0.09 & 0.100 & 0.981 & 1.39 & 0.908 & 1.29 &  0.13 & 0.084 & 0.952 & 0.99 & 0.960 & 0.96\\
                                               &                       & $\lambda_2$ &  0.01 & 0.064 & 0.934 & 1.05 & 0.948 & 1.02 & -0.03 & 0.040 & 0.952 & 0.79 & 0.940 & 0.77\\
			                                      &                       & $\beta$     & -0.03 & 0.033 & 0.893 & 0.71 & 0.934 & 0.62 & -0.09 & 0.018 & 0.987 & 0.52 & 0.967 & 0.49\\\cmidrule(lr){2-15}
			                                      & \multirow{4}{*}{0.75} & $\alpha$    &  0.04 & 0.052 & 0.996 & 0.95 & 0.914 & 0.38 &  0.02 & 0.004 & 0.944 & 0.25 & 0.945 & 0.24\\
			                                      &                       & $\lambda_1$ &  0.07 & 0.103 & 0.984 & 1.44 & 0.926 & 1.29 &  0.13 & 0.087 & 0.945 & 0.99 & 0.956 & 0.96\\
                                               &                       & $\lambda_2$ &  0.01 & 0.073 & 0.948 & 1.11 & 0.952 & 1.01 & -0.03 & 0.042 & 0.951 & 0.79 & 0.937 & 0.77\\
			                                      &                       & $\beta$     & -0.02 & 0.033 & 0.911 & 0.71 & 0.945 & 0.62 & -0.09 & 0.018 & 0.984 & 0.52 & 0.968 & 0.49\\\midrule
			\multirow{8}{*}{(0,...,10,...,0)} & \multirow{4}{*}{0.25}  & $\alpha$       &  0.03 & 0.012 & 0.986 & 0.46 & 0.887 & 0.36 &  0.01 & 0.004 & 0.946 & 0.24 & 0.946 & 0.24\\
			                                      &                       & $\lambda_1$ &  0.10 & 0.103 & 0.990 & 1.46 & 0.893 & 1.38 &  0.15 & 0.101 & 0.950 & 1.04 & 0.957 & 1.01\\
                                               &                       & $\lambda_2$ &  0.04 & 0.071 & 0.941 & 1.11 & 0.947 & 1.08 & -0.03 & 0.045 & 0.947 & 0.83 & 0.937 & 0.81\\
			                                      &                       & $\beta$     & -0.02 & 0.033 & 0.907 & 0.71 & 0.939 & 0.62 & -0.09 & 0.019 & 0.983 & 0.52 & 0.964 & 0.49\\\cmidrule(lr){2-15}
                                               & \multirow{4}{*}{0.75} & $\alpha$    &  0.04 & 0.017 & 0.983 & 0.47 & 0.877 & 0.37 &  0.02 & 0.004 & 0.949 & 0.24 & 0.949 & 0.24\\
			                                      &                       & $\lambda_1$ &  0.11 & 0.112 & 0.984 & 1.46 & 0.881 & 1.38 &  0.14 & 0.101 & 0.947 & 1.04 & 0.957 & 1.01\\
                                               &                       & $\lambda_2$ &  0.04 & 0.073 & 0.933 & 1.12 & 0.949 & 1.08 & -0.02 & 0.045 & 0.953 & 0.83 & 0.940 & 0.81\\
			                                      &                       & $\beta$     & -0.02 & 0.032 & 0.899 & 0.71 & 0.935 & 0.62 & -0.09 & 0.018 & 0.988 & 0.52 & 0.971 & 0.49\\
			\bottomrule
		\end{tabular}
	\end{center}
\end{table}  

\begin{table}[p]\tiny
    \caption{Performance of MLEs, Bayes estimates, confidence intervals, and credible
    intervals for $(\alpha, \lambda_1, \beta)$ = (1.5, 1.4, 1/1.4) under order
    restriction.}
   \label{tab:order3}
	\begin{center}
	\begin{tabular}{*{15}{c}}
		\toprule
		\multicolumn{3}{c}{} & \multicolumn{6}{c}{Likelihood Based} & \multicolumn{6}{c}{Bayesian}\\
		\cmidrule(lr){4-9}\cmidrule(lr){10-15}
         $\boldsymbol R$ & $T$ & Par & Bias & MSE & CPB & ALB & CPP & ALP & Bias & MSE & CPS & ALS & CPH & ALH \\ 
			\midrule
			\multirow{8}{*}{(0,...,0,10)}         & \multirow{4}{*}{0.25} & $\alpha$    & 0.07 & 0.057 & 0.958 & 0.90 & 0.918 & 0.89 &  0.06 & 0.051 & 0.946 & 0.83 & 0.945 & 0.82\\
			                                      &                       & $\lambda_1$ & 0.11 & 0.167 & 0.976 & 1.77 & 0.917 & 1.71 &  0.15 & 0.156 & 0.957 & 1.36 & 0.962 & 1.31\\
                                               &                       & $\lambda_2$ & 0.06 & 0.099 & 0.945 & 1.27 & 0.967 & 1.25 &  0.03 & 0.069 & 0.964 & 1.01 & 0.957 & 0.99\\
			                                      &                       & $\beta$     & 0.00 & 0.036 & 0.929 & 0.71 & 0.964 & 0.63 & -0.02 & 0.014 & 0.988 & 0.56 & 0.971 & 0.53\\\cmidrule(lr){2-15}
			                                      & \multirow{4}{*}{0.75} & $\alpha$    & 0.06 & 0.052 & 0.964 & 0.90 & 0.919 & 0.89 &  0.06 & 0.052 & 0.946 & 0.83 & 0.946 & 0.82\\
			                                      &                       & $\lambda_1$ & 0.13 & 0.182 & 0.985 & 1.81 & 0.897 & 1.75 &  0.16 & 0.175 & 0.949 & 1.38 & 0.959 & 1.33\\
                                               &                       & $\lambda_2$ & 0.05 & 0.092 & 0.944 & 1.27 & 0.969 & 1.24 &  0.04 & 0.076 & 0.959 & 1.02 & 0.955 & 1.00\\
			                                      &                       & $\beta$     & 0.01 & 0.038 & 0.925 & 0.70 & 0.962 & 0.63 & -0.02 & 0.014 & 0.988 & 0.56 & 0.972 & 0.53\\\midrule
			\multirow{8}{*}{(10,0,...,0)}         & \multirow{4}{*}{0.25} & $\alpha$    & 0.14 & 0.504 & 0.980 & 2.72 & 0.913 & 1.12 &  0.05 & 0.039 & 0.946 & 0.73 & 0.944 & 0.72\\
			                                      &                       & $\lambda_1$ & 0.07 & 0.136 & 0.971 & 1.58 & 0.924 & 1.47 &  0.11 & 0.105 & 0.956 & 1.16 & 0.956 & 1.12\\
                                               &                       & $\lambda_2$ & 0.03 & 0.082 & 0.939 & 1.15 & 0.955 & 1.11 &  0.01 & 0.052 & 0.955 & 0.88 & 0.948 & 0.86\\
			                                      &                       & $\beta$     & 0.01 & 0.038 & 0.920 & 0.70 & 0.962 & 0.63 & -0.02 & 0.014 & 0.987 & 0.56 & 0.974 & 0.53\\\cmidrule(lr){2-15}
			                                      & \multirow{4}{*}{0.75} & $\alpha$    & 0.14 & 0.484 & 0.993 & 2.83 & 0.905 & 1.12 &  0.05 & 0.040 & 0.950 & 0.73 & 0.949 & 0.72\\
			                                      &                       & $\lambda_1$ & 0.08 & 0.148 & 0.979 & 1.63 & 0.916 & 1.49 &  0.11 & 0.107 & 0.948 & 1.16 & 0.952 & 1.12\\
                                               &                       & $\lambda_2$ & 0.03 & 0.079 & 0.948 & 1.15 & 0.957 & 1.11 &  0.01 & 0.052 & 0.952 & 0.88 & 0.946 & 0.86\\
			                                      &                       & $\beta$     &-0.00 & 0.037 & 0.924 & 0.70 & 0.967 & 0.63 & -0.02 & 0.014 & 0.988 & 0.56 & 0.974 & 0.53\\\midrule
			\multirow{8}{*}{(0,...,10,...,0)}     & \multirow{4}{*}{0.25} & $\alpha$    & 0.06 & 0.049 & 0.975 & 0.93 & 0.923 & 0.90 &  0.06 & 0.051 & 0.947 & 0.83 & 0.946 & 0.82\\
			                                      &                       & $\lambda_1$ & 0.11 & 0.163 & 0.978 & 1.81 & 0.913 & 1.74 &  0.14 & 0.166 & 0.954 & 1.35 & 0.962 & 1.31\\
                                               &                       & $\lambda_2$ & 0.07 & 0.109 & 0.956 & 1.30 & 0.964 & 1.27 &  0.04 & 0.074 & 0.961 & 1.01 & 0.953 & 0.99\\
			                                      &                       & $\beta$     & 0.01 & 0.037 & 0.923 & 0.71 & 0.960 & 0.63 & -0.02 & 0.014 & 0.990 & 0.56 & 0.979 & 0.53\\\cmidrule(lr){2-15}
			                                      & \multirow{4}{*}{0.75} & $\alpha$    & 0.09 & 0.109 & 0.989 & 1.37 & 0.904 & 1.09 &  0.05 & 0.040 & 0.946 & 0.73 & 0.946 & 0.72\\
			                                      &                       & $\lambda_1$ & 0.10 & 0.144 & 0.980 & 1.68 & 0.923 & 1.59 &  0.12 & 0.131 & 0.952 & 1.23 & 0.957 & 1.19\\
                                               &                       & $\lambda_2$ & 0.06 & 0.098 & 0.952 & 1.25 & 0.958 & 1.20 &  0.02 & 0.062 & 0.956 & 0.93 & 0.947 & 0.90\\
			                                      &                       & $\beta$     & 0.00 & 0.038 & 0.928 & 0.70 & 0.960 & 0.63 & -0.02 & 0.015 & 0.985 & 0.56 & 0.971 & 0.53\\
			\bottomrule
		\end{tabular}
	\end{center}
\end{table}  

\begin{table}[p]\tiny
    \caption{Performance of MLEs, Bayes estimates, confidence intervals, and credible
    intervals for $(\alpha, \lambda_1, \beta)$ = (0.5, 1.4, 1/1.4) under order
    restriction.}
   \label{tab:order4}
	\begin{center}
		\begin{tabular}{*{15}{c}}
			\toprule
            \multicolumn{3}{c}{} & \multicolumn{6}{c}{Likelihood Based} & \multicolumn{6}{c}{Bayesian}\\
            \cmidrule(lr){4-9}\cmidrule(lr){10-15}
         $\boldsymbol R$ & $T$ & Par & Bias & MSE & CPB & ALB & CPP & ALP & Bias & MSE & CPS & ALS & CPH & ALH \\ 
			\midrule
         \multirow{8}{*}{(0,...,0,10)}         & \multirow{4}{*}{0.25} & $\alpha$    & 0.02 & 0.006 & 0.968 & 0.30 & 0.920 & 0.30 &  0.02 & 0.006 & 0.947 & 0.28 & 0.947 & 0.27\\
			                                      &                       & $\lambda_1$ & 0.11 & 0.149 & 0.981 & 1.78 & 0.922 & 1.71 &  0.15 & 0.166 & 0.954 & 1.37 & 0.962 & 1.31\\
                                               &                       & $\lambda_2$ & 0.06 & 0.086 & 0.956 & 1.27 & 0.965 & 1.25 &  0.04 & 0.072 & 0.963 & 1.02 & 0.961 & 0.99\\
			                                      &                       & $\beta$     & 0.01 & 0.038 & 0.919 & 0.70 & 0.956 & 0.63 & -0.02 & 0.014 & 0.989 & 0.56 & 0.977 & 0.53\\\cmidrule(lr){2-15}
			                                      & \multirow{4}{*}{0.75} & $\alpha$    & 0.02 & 0.006 & 0.952 & 0.30 & 0.921 & 0.30 &  0.02 & 0.006 & 0.947 & 0.28 & 0.947 & 0.27\\
			                                      &                       & $\lambda_1$ & 0.11 & 0.163 & 0.980 & 1.78 & 0.902 & 1.71 &  0.15 & 0.160 & 0.956 & 1.37 & 0.959 & 1.31\\
                                               &                       & $\lambda_2$ & 0.06 & 0.102 & 0.949 & 1.27 & 0.956 & 1.25 &  0.04 & 0.072 & 0.960 & 1.02 & 0.956 & 0.99\\
			                                      &                       & $\beta$     & 0.01 & 0.038 & 0.922 & 0.70 & 0.959 & 0.63 & -0.02 & 0.014 & 0.989 & 0.56 & 0.975 & 0.53\\\midrule
			\multirow{8}{*}{(10,0,...,0)}         & \multirow{4}{*}{0.25} & $\alpha$    & 0.05 & 0.054 & 0.989 & 0.93 & 0.904 & 0.38 &  0.01 & 0.004 & 0.948 & 0.24 & 0.948 & 0.24\\
			                                      &                       & $\lambda_1$ & 0.07 & 0.139 & 0.977 & 1.74 & 0.919 & 1.47 &  0.11 & 0.099 & 0.961 & 1.16 & 0.965 & 1.12\\
                                               &                       & $\lambda_2$ & 0.03 & 0.084 & 0.938 & 1.26 & 0.947 & 1.11 &  0.01 & 0.048 & 0.970 & 0.89 & 0.962 & 0.87\\
			                                      &                       & $\beta$     & 0.01 & 0.038 & 0.919 & 0.70 & 0.954 & 0.63 & -0.02 & 0.014 & 0.990 & 0.56 & 0.975 & 0.53\\\cmidrule(lr){2-15}
			                                      & \multirow{4}{*}{0.75} & $\alpha$    & 0.05 & 0.044 & 0.995 & 0.96 & 0.892 & 0.37 &  0.02 & 0.004 & 0.952 & 0.25 & 0.953 & 0.24\\
			                                      &                       & $\lambda_1$ & 0.07 & 0.134 & 0.973 & 1.61 & 0.922 & 1.48 &  0.11 & 0.108 & 0.956 & 1.16 & 0.956 & 1.13\\
                                               &                       & $\lambda_2$ & 0.04 & 0.078 & 0.953 & 1.17 & 0.965 & 1.12 &  0.01 & 0.051 & 0.961 & 0.89 & 0.953 & 0.87\\
			                                      &                       & $\beta$     & 0.01 & 0.036 & 0.930 & 0.71 & 0.969 & 0.63 & -0.02 & 0.014 & 0.991 & 0.56 & 0.977 & 0.53\\\midrule
			\multirow{8}{*}{(0,...,10,...,0)} & \multirow{4}{*}{0.25}     & $\alpha$    & 0.03 & 0.011 & 0.987 & 0.46 & 0.906 & 0.36 &  0.02 & 0.004 & 0.946 & 0.24 & 0.946 & 0.24\\
			                                      &                       & $\lambda_1$ & 0.10 & 0.142 & 0.974 & 1.72 & 0.909 & 1.61 &  0.12 & 0.128 & 0.952 & 1.22 & 0.954 & 1.18\\
                                               &                       & $\lambda_2$ & 0.04 & 0.080 & 0.949 & 1.23 & 0.965 & 1.18 &  0.02 & 0.059 & 0.956 & 0.93 & 0.952 & 0.90\\
			                                      &                       & $\beta$     & 0.00 & 0.037 & 0.917 & 0.70 & 0.957 & 0.63 & -0.02 & 0.014 & 0.989 & 0.55 & 0.975 & 0.53\\\cmidrule(lr){2-15}
                                               & \multirow{4}{*}{0.75} & $\alpha$    & 0.03 & 0.012 & 0.989 & 0.46 & 0.897 & 0.36 &  0.02 & 0.004 & 0.954 & 0.24 & 0.951 & 0.24\\
			                                      &                       & $\lambda_1$ & 0.11 & 0.142 & 0.978 & 1.71 & 0.897 & 1.60 &  0.12 & 0.123 & 0.953 & 1.22 & 0.958 & 1.18\\
                                               &                       & $\lambda_2$ & 0.05 & 0.081 & 0.944 & 1.23 & 0.958 & 1.18 &  0.02 & 0.058 & 0.959 & 0.93 & 0.955 & 0.90\\
			                                      &                       & $\beta$     & 0.00 & 0.037 & 0.921 & 0.70 & 0.952 & 0.63 & -0.02 & 0.014 & 0.988 & 0.56 & 0.976 & 0.53\\
			\bottomrule
		\end{tabular}
	\end{center}
\end{table}  

\begin{table}\tiny
    \caption{Performance of MLEs, Bayes estimates, confidence intervals, and credible
    intervals for $(\alpha, \lambda_1, \lambda_2)$ = (1.5, 1.2, 1) under unrestricted case.}
   \label{tab:unorder1}
   \begin{center}
      \begin{tabular}{*{15}{c}}
         \toprule
         & & & \multicolumn{6}{c}{Likelihood Based} & \multicolumn{6}{c}{Bayesian}\\ 
         \cmidrule(lr){4-9}\cmidrule(lr){10-15}
         $\boldsymbol R$ & $T$ & Par & Bias & MSE & CPB & ALB & CPP & ALP & Bias & MSE & CPS & ALS & CPH & ALH \\ 
         \midrule
         \multirow{6}{*}{(0,...,0,10)}         & \multirow{3}{*}{0.25} & $\alpha$    & 0.06 & 0.051 & 0.964 & 0.90 & 0.914 & 0.89 & 0.06 & 0.054 & 0.943 & 0.83 & 0.946 & 0.83\\
                                               &                       & $\lambda_1$ & 0.08 & 0.119 & 0.965 & 1.47 & 0.940 & 1.43 & 0.09 & 0.131 & 0.944 & 1.23 & 0.948 & 1.19\\
                                               &                       & $\lambda_2$ & 0.06 & 0.094 & 0.965 & 1.31 & 0.936 & 1.28 & 0.08 & 0.106 & 0.938 & 1.10 & 0.938 & 1.07\\\cmidrule(lr){2-15}
                                               & \multirow{3}{*}{0.75} & $\alpha$    & 0.06 & 0.049 & 0.966 & 0.90 & 0.932 & 0.89 & 0.05 & 0.049 & 0.950 & 0.83 & 0.948 & 0.82\\
                                               &                       & $\lambda_1$ & 0.09 & 0.115 & 0.963 & 1.50 & 0.934 & 1.48 & 0.08 & 0.118 & 0.943 & 1.22 & 0.946 & 1.18\\
                                               &                       & $\lambda_2$ & 0.08 & 0.094 & 0.964 & 1.33 & 0.922 & 1.30 & 0.06 & 0.093 & 0.946 & 1.09 & 0.944 & 1.05\\ \midrule
         \multirow{6}{*}{(10,0,...,0)}         & \multirow{3}{*}{0.25} & $\alpha$    & 0.10 & 0.339 & 0.990 & 2.76 & 0.919 & 1.11 & 0.05 & 0.040 & 0.947 & 0.73 & 0.944 & 0.72\\
                                               &                       & $\lambda_1$ & 0.06 & 0.112 & 0.961 & 1.30 & 0.937 & 1.26 & 0.05 & 0.092 & 0.941 & 1.08 & 0.937 & 1.05\\
                                               &                       & $\lambda_2$ & 0.04 & 0.085 & 0.954 & 1.16 & 0.934 & 1.13 & 0.04 & 0.076 & 0.940 & 0.98 & 0.932 & 0.95\\ \cmidrule(lr){2-15}
                                               & \multirow{3}{*}{0.75} & $\alpha$    & 0.12 & 0.292 & 0.996 & 2.84 & 0.898 & 1.16 & 0.04 & 0.040 & 0.941 & 0.73 & 0.939 & 0.72\\
                                               &                       & $\lambda_1$ & 0.06 & 0.095 & 0.966 & 1.33 & 0.945 & 1.27 & 0.04 & 0.089 & 0.945 & 1.07 & 0.938 & 1.04\\
                                               &                       & $\lambda_2$ & 0.05 & 0.083 & 0.959 & 1.18 & 0.937 & 1.14 & 0.05 & 0.074 & 0.940 & 0.98 & 0.937 & 0.95\\ \midrule
         \multirow{6}{*}{(0,...,10,...,0)}     & \multirow{3}{*}{0.25} & $\alpha$    & 0.06 & 0.057 & 0.963 & 0.92 & 0.909 & 0.90 & 0.06 & 0.053 & 0.947 & 0.83 & 0.948 & 0.83\\
                                               &                       & $\lambda_1$ & 0.11 & 0.134 & 0.976 & 1.49 & 0.928 & 1.48 & 0.09 & 0.126 & 0.949 & 1.23 & 0.952 & 1.20\\
                                               &                       & $\lambda_2$ & 0.07 & 0.107 & 0.958 & 1.31 & 0.916 & 1.28 & 0.07 & 0.099 & 0.943 & 1.10 & 0.940 & 1.07\\\cmidrule(lr){2-15}
                                               & \multirow{3}{*}{0.75} & $\alpha$    & 0.08 & 0.121 & 0.988 & 1.38 & 0.885 & 1.05 & 0.05 & 0.040 & 0.949 & 0.73 & 0.947 & 0.72\\
                                               &                       & $\lambda_1$ & 0.06 & 0.106 & 0.959 & 1.42 & 0.940 & 1.37 & 0.06 & 0.099 & 0.944 & 1.12 & 0.942 & 1.09\\
                                               &                       & $\lambda_2$ & 0.08 & 0.094 & 0.968 & 1.25 & 0.932 & 1.21 & 0.05 & 0.082 & 0.943 & 1.01 & 0.941 & 0.98\\
         \bottomrule
      \end{tabular}
   \end{center}
\end{table}  

\begin{table}[p]\tiny
    \caption{Performance of MLEs, Bayes estimates, confidence intervals, and credible
    intervals for $(\alpha, \lambda_1, \lambda_2)$ = (0.5, 1.2, 1) under unrestricted case.}
   \label{tab:unorder2}
\begin{center}
	\begin{tabular}{*{15}{c}}
		\toprule
      & & & \multicolumn{6}{c}{Likelihood Based} & \multicolumn{6}{c}{Bayesian}\\ 
      \cmidrule(lr){4-9}\cmidrule(lr){10-15}
      $\boldsymbol R$ & $T$ & Par & Bias & MSE & CPB & ALB & CPP & ALP & Bias & MSE & CPS & ALS & CPH & ALH \\ 
		\midrule
		\multirow{6}{*}{(0,...,0,10)}         & \multirow{3}{*}{0.25} & $\alpha$    & 0.02 & 0.006 & 0.965 & 0.30 & 0.920 & 0.29 & 0.02 & 0.006 & 0.941 & 0.28 & 0.940 & 0.28\\
		                                      &                       & $\lambda_1$ & 0.07 & 0.114 & 0.965 & 1.45 & 0.931 & 1.42 & 0.09 & 0.122 & 0.946 & 1.23 & 0.949 & 1.19\\
		                                      &                       & $\lambda_2$ & 0.07 & 0.092 & 0.966 & 1.30 & 0.931 & 1.27 & 0.07 & 0.100 & 0.945 & 1.10 & 0.940 & 1.06\\\cmidrule(lr){2-15}
		                                      & \multirow{3}{*}{0.75} & $\alpha$    & 0.02 & 0.006 & 0.956 & 0.30 & 0.923 & 0.30 & 0.02 & 0.006 & 0.949 & 0.28 & 0.951 & 0.27\\
		                                      &                       & $\lambda_1$ & 0.09 & 0.118 & 0.965 & 1.49 & 0.937 & 1.46 & 0.08 & 0.117 & 0.952 & 1.22 & 0.948 & 1.18\\
		                                      &                       & $\lambda_2$ & 0.07 & 0.090 & 0.964 & 1.32 & 0.937 & 1.29 & 0.07 & 0.091 & 0.948 & 1.09 & 0.945 & 1.05\\\midrule
		\multirow{6}{*}{(10,0,...,0)}         & \multirow{3}{*}{0.25} & $\alpha$    & 0.05 & 0.083 & 0.986 & 0.93 & 0.906 & 0.38 & 0.01 & 0.004 & 0.949 & 0.24 & 0.948 & 0.24\\
		                                      &                       & $\lambda_1$ & 0.05 & 0.115 & 0.953 & 1.34 & 0.940 & 1.29 & 0.05 & 0.091 & 0.946 & 1.08 & 0.939 & 1.06\\
		                                      &                       & $\lambda_2$ & 0.04 & 0.092 & 0.954 & 1.19 & 0.939 & 1.16 & 0.04 & 0.072 & 0.943 & 0.98 & 0.942 & 0.96\\\cmidrule(lr){2-15}
		                                      & \multirow{3}{*}{0.75} & $\alpha$    & 0.06 & 0.107 & 0.990 & 0.95 & 0.899 & 0.38 & 0.01 & 0.004 & 0.948 & 0.25 & 0.950 & 0.24\\
		                                      &                       & $\lambda_1$ & 0.05 & 0.118 & 0.960 & 1.33 & 0.929 & 1.28 & 0.04 & 0.085 & 0.946 & 1.07 & 0.941 & 1.05\\
		                                      &                       & $\lambda_2$ & 0.03 & 0.093 & 0.942 & 1.17 & 0.929 & 1.14 & 0.05 & 0.070 & 0.946 & 0.98 & 0.943 & 0.96\\\midrule
      \multirow{6}{*}{(0,...,10,...,0)} & \multirow{3}{*}{0.25}     & $\alpha$    & 0.03 & 0.013 & 0.982 & 0.46 & 0.906 & 0.36 & 0.01 & 0.004 & 0.947 & 0.24 & 0.950 & 0.24\\
		                                      &                       & $\lambda_1$ & 0.06 & 0.108 & 0.953 & 1.40 & 0.929 & 1.35 & 0.06 & 0.099 & 0.946 & 1.12 & 0.943 & 1.09\\
		                                      &                       & $\lambda_2$ & 0.05 & 0.086 & 0.954 & 1.24 & 0.934 & 1.20 & 0.05 & 0.076 & 0.945 & 1.01 & 0.939 & 0.98\\\cmidrule(lr){2-15}
		                                      & \multirow{3}{*}{0.75} & $\alpha$    & 0.03 & 0.014 & 0.983 & 0.46 & 0.910 & 0.36 & 0.02 & 0.004 & 0.948 & 0.24 & 0.949 & 0.24\\
		                                      &                       & $\lambda_1$ & 0.05 & 0.097 & 0.957 & 1.41 & 0.938 & 1.36 & 0.06 & 0.100 & 0.949 & 1.13 & 0.945 & 1.10\\
		                                      &                       & $\lambda_2$ & 0.05 & 0.081 & 0.953 & 1.26 & 0.931 & 1.23 & 0.06 & 0.083 & 0.943 & 1.02 & 0.938 & 0.99\\
		\bottomrule
	\end{tabular}
\end{center}
\end{table}  

\begin{table}[p]\tiny
    \caption{Performance of MLEs, Bayes estimates, confidence intervals, and credible
    intervals for $(\alpha, \lambda_1, \lambda_2)$ = (1.5, 1.4, 1) under unrestricted case.}
   \label{tab:unorder3}
\begin{center}
	\begin{tabular}{*{15}{c}}
		\toprule
      & & & \multicolumn{6}{c}{Likelihood Based} & \multicolumn{6}{c}{Bayesian}\\ 
      \cmidrule(lr){4-9}\cmidrule(lr){10-15}
      $\boldsymbol R$ & $T$ & Par & Bias & MSE & CPB & ALB & CPP & ALP & Bias & MSE & CPS & ALS & CPH & ALH \\ 
		\midrule
      \multirow{6}{*}{(0,...,0,10)}         & \multirow{3}{*}{0.25} & $\alpha$    & 0.06 & 0.053 & 0.968 & 0.90 & 0.919 & 0.89 & 0.055 & 0.052 & 0.947 & 0.83 & 0.950 & 0.82\\
		                                      &                       & $\lambda_1$ & 0.11 & 0.186 & 0.965 & 1.78 & 0.913 & 1.74 & 0.116 & 0.180 & 0.944 & 1.45 & 0.945 & 1.41\\
		                                      &                       & $\lambda_2$ & 0.09 & 0.115 & 0.969 & 1.43 & 0.935 & 1.40 & 0.082 & 0.111 & 0.945 & 1.17 & 0.947 & 1.13\\\cmidrule(lr){2-15}
		                                      & \multirow{3}{*}{0.75} & $\alpha$    & 0.05 & 0.047 & 0.967 & 0.90 & 0.930 & 0.89 & 0.055 & 0.052 & 0.948 & 0.83 & 0.946 & 0.82\\
		                                      &                       & $\lambda_1$ & 0.08 & 0.148 & 0.964 & 1.73 & 0.934 & 1.68 & 0.111 & 0.183 & 0.947 & 1.44 & 0.949 & 1.40\\
		                                      &                       & $\lambda_2$ & 0.07 & 0.101 & 0.960 & 1.38 & 0.933 & 1.35 & 0.081 & 0.115 & 0.948 & 1.17 & 0.948 & 1.13\\\midrule
		\multirow{6}{*}{(10,0,...,0)}         & \multirow{3}{*}{0.25} & $\alpha$    & 0.13 & 0.455 & 0.984 & 2.73 & 0.916 & 1.10 & 0.040 & 0.039 & 0.945 & 0.73 & 0.944 & 0.72\\
		                                      &                       & $\lambda_1$ & 0.06 & 0.138 & 0.967 & 1.59 & 0.933 & 1.50 & 0.069 & 0.123 & 0.944 & 1.25 & 0.945 & 1.21\\
		                                      &                       & $\lambda_2$ & 0.04 & 0.095 & 0.949 & 1.27 & 0.937 & 1.22 & 0.044 & 0.084 & 0.936 & 1.03 & 0.932 & 1.00\\\cmidrule(lr){2-15}
		                                      & \multirow{3}{*}{0.75} & $\alpha$    & 0.12 & 0.408 & 0.996 & 2.83 & 0.899 & 1.11 & 0.043 & 0.040 & 0.942 & 0.73 & 0.943 & 0.72\\
		                                      &                       & $\lambda_1$ & 0.07 & 0.137 & 0.969 & 1.59 & 0.940 & 1.50 & 0.064 & 0.120 & 0.941 & 1.24 & 0.941 & 1.21\\
		                                      &                       & $\lambda_2$ & 0.04 & 0.083 & 0.962 & 1.26 & 0.951 & 1.21 & 0.047 & 0.083 & 0.940 & 1.03 & 0.935 & 1.00\\\midrule
      \multirow{6}{*}{(0,...,10,...,0)} & \multirow{3}{*}{0.25}     & $\alpha$    & 0.06 & 0.049 & 0.982 & 0.92 & 0.929 & 0.89 & 0.060 & 0.054 & 0.942 & 0.83 & 0.944 & 0.83\\
		                                      &                       & $\lambda_1$ & 0.09 & 0.168 & 0.971 & 1.75 & 0.936 & 1.70 & 0.114 & 0.181 & 0.946 & 1.45 & 0.947 & 1.40\\
		                                      &                       & $\lambda_2$ & 0.07 & 0.104 & 0.959 & 1.40 & 0.938 & 1.36 & 0.086 & 0.120 & 0.941 & 1.17 & 0.941 & 1.13\\\cmidrule(lr){2-15}
		                                      & \multirow{3}{*}{0.75} & $\alpha$    & 0.10 & 0.118 & 0.985 & 1.40 & 0.886 & 1.11 & 0.047 & 0.039 & 0.946 & 0.73 & 0.948 & 0.72\\
		                                      &                       & $\lambda_1$ & 0.10 & 0.146 & 0.969 & 1.71 & 0.929 & 1.63 & 0.079 & 0.138 & 0.948 & 1.30 & 0.946 & 1.27\\
		                                      &                       & $\lambda_2$ & 0.07 & 0.088 & 0.968 & 1.36 & 0.941 & 1.30 & 0.059 & 0.089 & 0.946 & 1.08 & 0.944 & 1.04\\
		\bottomrule
	\end{tabular}
\end{center}
\end{table}  

\begin{table}[p]\tiny
    \caption{Performance of MLEs, Bayes estimates, confidence intervals, and credible
    intervals for $(\alpha, \lambda_1, \lambda_2)$ = (0.5, 1.4, 1) under unrestricted case.}
   \label{tab:unorder4}
\begin{center}
	\begin{tabular}{*{15}{c}}
		\toprule
      & & & \multicolumn{6}{c}{Likelihood Based} & \multicolumn{6}{c}{Bayesian}\\ 
      \cmidrule(lr){4-9}\cmidrule(lr){10-15}
      $\boldsymbol R$ & $T$ & Par & Bias & MSE & CPB & ALB & CPP & ALP & Bias & MSE & CPS & ALS & CPH & ALH \\ 
		\midrule
		\multirow{6}{*}{(0,...,0,10)}         & \multirow{3}{*}{0.25} & $\alpha$    & 0.02 & 0.006 & 0.966 & 0.30 & 0.927 & 0.30 & 0.021 & 0.006 & 0.949 & 0.28 & 0.952 & 0.28\\
		                                      &                       & $\lambda_1$ & 0.11 & 0.175 & 0.968 & 1.78 & 0.918 & 1.73 & 0.117 & 0.175 & 0.949 & 1.45 & 0.949 & 1.40\\
		                                      &                       & $\lambda_2$ & 0.07 & 0.109 & 0.954 & 1.41 & 0.942 & 1.38 & 0.078 & 0.109 & 0.949 & 1.17 & 0.944 & 1.12\\\cmidrule(lr){2-15}
		                                      & \multirow{3}{*}{0.75} & $\alpha$    & 0.02 & 0.006 & 0.962 & 0.30 & 0.909 & 0.30 & 0.020 & 0.006 & 0.949 & 0.28 & 0.949 & 0.28\\
		                                      &                       & $\lambda_1$ & 0.11 & 0.159 & 0.964 & 1.78 & 0.937 & 1.73 & 0.114 & 0.178 & 0.941 & 1.44 & 0.944 & 1.40\\
		                                      &                       & $\lambda_2$ & 0.08 & 0.115 & 0.957 & 1.42 & 0.933 & 1.38 & 0.077 & 0.108 & 0.948 & 1.17 & 0.944 & 1.12\\\midrule
		\multirow{6}{*}{(10,0,...,0)}         & \multirow{3}{*}{0.25} & $\alpha$    & 0.04 & 0.028 & 0.995 & 0.92 & 0.904 & 0.37 & 0.017 & 0.005 & 0.942 & 0.25 & 0.945 & 0.24\\
		                                      &                       & $\lambda_1$ & 0.06 & 0.131 & 0.964 & 1.60 & 0.938 & 1.50 & 0.068 & 0.122 & 0.949 & 1.25 & 0.945 & 1.22\\
		                                      &                       & $\lambda_2$ & 0.05 & 0.099 & 0.953 & 1.28 & 0.926 & 1.23 & 0.046 & 0.080 & 0.947 & 1.04 & 0.941 & 1.01\\\cmidrule(lr){2-15}
		                                      & \multirow{3}{*}{0.75} & $\alpha$    & 0.05 & 0.055 & 0.993 & 0.95 & 0.903 & 0.38 & 0.016 & 0.004 & 0.947 & 0.25 & 0.948 & 0.24\\
		                                      &                       & $\lambda_1$ & 0.06 & 0.140 & 0.955 & 1.56 & 0.936 & 1.49 & 0.061 & 0.120 & 0.941 & 1.25 & 0.936 & 1.22\\
		                                      &                       & $\lambda_2$ & 0.03 & 0.091 & 0.944 & 1.25 & 0.936 & 1.21 & 0.048 & 0.080 & 0.945 & 1.04 & 0.939 & 1.01\\\midrule
      \multirow{6}{*}{(0,...,10,...,0)} & \multirow{3}{*}{0.25}     & $\alpha$    & 0.03 & 0.012 & 0.986 & 0.46 & 0.890 & 0.37 & 0.016 & 0.004 & 0.946 & 0.24 & 0.948 & 0.24\\
		                                      &                       & $\lambda_1$ & 0.10 & 0.160 & 0.965 & 1.72 & 0.928 & 1.64 & 0.071 & 0.133 & 0.947 & 1.30 & 0.943 & 1.26\\
		                                      &                       & $\lambda_2$ & 0.07 & 0.092 & 0.969 & 1.37 & 0.939 & 1.31 & 0.055 & 0.086 & 0.947 & 1.07 & 0.943 & 1.04\\\cmidrule(lr){2-15}
		                                      & \multirow{3}{*}{0.75} & $\alpha$    & 0.03 & 0.018 & 0.984 & 0.47 & 0.906 & 0.37 & 0.015 & 0.005 & 0.937 & 0.24 & 0.938 & 0.24\\
		                                      &                       & $\lambda_1$ & 0.08 & 0.143 & 0.963 & 1.69 & 0.929 & 1.60 & 0.084 & 0.146 & 0.939 & 1.31 & 0.941 & 1.28\\
		                                      &                       & $\lambda_2$ & 0.06 & 0.090 & 0.956 & 1.34 & 0.935 & 1.28 & 0.056 & 0.090 & 0.948 & 1.08 & 0.943 & 1.04\\
		\bottomrule
	\end{tabular}
\end{center}
\end{table}  

\begin{table}[ht]\scriptsize
    \caption{\scriptsize MLEs and 95\% confidence intervals constructed using parametric
    bootstrap (BB), percentile bootstrap (PB), and Bayes estimates and 95\% symmetric
    (SCRI) and HPD credible interval (HPD CRI) for model parameters based on the simulated
    dataset when there is order restriction on parameters.}
    \label{tab:DataAnalysisOrder}
	\begin{center}
		\begin{tabular}{*{7}{c}}
			\toprule
            {} & \multicolumn{3}{c}{Likelihood} & \multicolumn{3}{c}{Bayesian}\\
            \cmidrule(lr){2-4}\cmidrule(lr){5-7}
			Parameter & MLE & BB & PB & BE & SCRI & HPD CRI \\ 
			\midrule
			$\alpha$    & 1.49 & (1.22, 1.72) & (1.30, 1.81) & 1.49 & (1.26, 1.75) & (1.27, 1.76)\\
			$\lambda_1$ & 1.62 & (1.11, 2.04) & (1.23, 2.16) & 1.58 & (1.19, 2.06) & (1.19, 2.04)\\
			$\lambda_2$ & 0.89 & (0.53, 1.20) & (0.62, 1.27) & 0.91 & (0.64, 1.26) & (0.61, 1.22)\\
			\bottomrule
		\end{tabular}
	\end{center}
\end{table}

\begin{table}[ht]\scriptsize
    \caption{\scriptsize MLEs and 95\% confidence intervals constructed using parametric
    bootstrap (BB), percentile bootstrap (PB), and Bayes estimates and 95\% symmetric
    (SCRI) and HPD credible interval (HPD CRI) for model parameters based on the simulated
    dataset when there is no order restriction on parameters.}
    \label{tab:DataAnalysisUnorder}
	\begin{center}
		\begin{tabular}{*{7}{c}}
			\toprule
            {} & \multicolumn{3}{c}{Likelihood} & \multicolumn{3}{c}{Bayesian}\\
            \cmidrule(lr){2-4}\cmidrule(lr){5-7}
			Parameter & MLE & BB & PB & BE & SCRI & HPD CRI \\ 
			\midrule
			$\alpha$    & 1.49 & (1.23, 1.73) & (1.286, 1.770) & 1.49 & (1.26, 1.74) & (1.25, 1.72)\\
			$\lambda_1$ & 1.62 & (1.13, 2.05) & (1.213, 2.125) & 1.61 & (1.20, 2.09) & (1.18, 2.05)\\
			$\lambda_2$ & 0.89 & (0.52, 1.20) & (0.605, 1.283) & 0.89 & (0.60, 1.24) & (0.60, 1.23)\\
            \bottomrule
		\end{tabular}
	\end{center}
\end{table}

\newpage

\section*{\sc Acknowledgements} 
The research of Ayon Ganguly is supported by the Mathematical Research Impact
Centric Support (File no.~MTR/2017/000700) from the Science and
Engineering Research Board, Department of Science and Technology, Government of
India. \\
Debanjan Mitra thanks Indian Institute of Management Udaipur for financial
support to carry out this research.

\singlespacing
\bibliographystyle{acm}
\bibliography{article.bib,book.bib}

\begin{thebibliography}{10}

\bibitem{BBK2009}
{\sc Balakrishnan, N., Beutner, E., and Kateri, M.}
\newblock Order restricted inference for exponential step-stress models.
\newblock {\em IEEE Transactions on Reliability 58\/} (2009), 132--142.

\bibitem{B:BC2014}
{\sc Balakrishnan, N., and Cramer, E.}
\newblock {\em The art of Progressive censoring: applications to reliability
  and quality}.
\newblock Birkh{\"{a}}user, Boston, 2014.

\bibitem{BS1993}
{\sc Berger, J.~O., and Sun, D.}
\newblock Bayesian analysis for the {Poly-Weibull} distribution.
\newblock {\em Journal of American Statistical Association 88\/} (1993),
  1412--1418.

\bibitem{B2008}
{\sc Burkschat, M.}
\newblock On optimality of extremal schemes in progressive type {II} censoring.
\newblock {\em Journal of Statistical Planning and Inference\/} (2008),
  1647--1659.

\bibitem{C1959}
{\sc Cox, D.~R.}
\newblock The analysis of exponentially distributed lifetimes with two types of
  failures.
\newblock {\em Journal of the Royal Statistical Society, Series B 21\/} (1959),
  411--421.

\bibitem{B:DN2003}
{\sc David, H.~A., and Nagaraja, H.~N.}
\newblock {\em Order statistics}, 3~ed.
\newblock John Wiley and Sons, New York, 2003.

\bibitem{K2004}
{\sc Kundu, D.}
\newblock Parameter estimation of the partially complete time and type of
  failure data.
\newblock {\em Biometrical Journal 46\/} (2004), 165--179.

\bibitem{K2008}
{\sc Kundu, D.}
\newblock Bayesian inference and life testing plan for {Weibull} distribution
  in presence of progressive censoring.
\newblock {\em Technometrics 50\/} (2008), 144--154.

\bibitem{KG2006}
{\sc Kundu, D., and Gupta, R.}
\newblock Estimation of {$P(Y < X)$} for {Weibull} distribution.
\newblock {\em IEEE Transactions on Reliability 55\/} (2006), 270--280.

\bibitem{MLTW2021}
{\sc Mahto, A., Lodhi, C., Tripathi, Y., and Wang, L.}
\newblock Inference for partially observed competing risks model for
  {Kumaraswamy} distribution under generalized progressive hybrid censoring.
\newblock {\em Journal of Applied Statistics\/} (2021).

\bibitem{NCB2004}
{\sc Ng, H. K.~T., Chan, P.~S., and Balakrishnan, N.}
\newblock Optimum progressive censoring plan for the {Weibull} distribution.
\newblock {\em Technometrics 46\/} (2004), 470--481.

\bibitem{NKC2009}
{\sc Ng, H. K.~T., Kundu, D., and Chan, P.~S.}
\newblock Statistical analysis of exponential lifetimes under an adaptive
  {Type-II} progressive censoring scheme.
\newblock {\em Naval Research Logistics 56\/} (2009), 687--698.

\bibitem{AMK2021b}
{\sc Pal, A., Mitra, S., and Kundu, D.}
\newblock Bayesian order restricted inference of a weibull multi-step
  step-stress model.
\newblock {\em Journal of Statistical Theory and Practice 15\/} (2021).

\bibitem{AMK2021a}
{\sc Pal, A., Mitra, S., and Kundu, D.}
\newblock Order restricted classical inference of a weibull multiple
  step-stress model.
\newblock {\em Journal of Applied Statistics 48\/} (2021), 623 --645.

\bibitem{PKK2009}
{\sc Pareek, B., Kundu, D., and Kumar, S.}
\newblock On progressively censored competing risks data for {Weibull}
  distributions.
\newblock {\em Computational Statistics and Data Analysis 53\/} (2009),
  4083--4094.

\bibitem{P2007}
{\sc Pascual, F.}
\newblock Accelerated life test planning with independent {Weibull} competing
  risks with known shape parameter.
\newblock {\em IEEE Transactions on Reliability 56\/} (2007), 85--93.

\bibitem{P2008}
{\sc Pascual, F.}
\newblock Accelerated life test planning with independent {Weibull} competing
  risks.
\newblock {\em IEEE Transactions on Reliability 57\/} (2008), 435--444.

\bibitem{PKPFFB1978}
{\sc Prentice, R.~L., Kalbfleish, J., Peterson, J.~A., Flurnoy, N., Farewell,
  V.~T., and Breslow, N.}
\newblock The analysis of failure time points in presence of competing risks.
\newblock {\em Biometrics 34\/} (1978), 541 -- 544.

\bibitem{RG2021}
{\sc Ren, J., and Gui, W.}
\newblock Statistical analysis of adaptive type-{II} progressively censored
  competing risks for {W}eibull moels.
\newblock {\em Applied Mathematical Modelling 98\/} (2021), 323--342.

\bibitem{SGKM2017}
{\sc Samanta, D., Ganguly, A., Kundu, D., and Mitra, S.}
\newblock Order restricted {Bayesian} inference for exponential simple
  step-stress model.
\newblock {\em Communication in Statistics - Simulation and Computation 46\/}
  (2017), 1113--1135.

\end{thebibliography}

\end{document}